\documentclass{aa}  

\usepackage{graphicx}
\usepackage[varg]{txfonts}
\usepackage{hyperref}
\newcommand{\jednad}{\mbox{1-D}}
\newcommand{\dvad}{\mbox{2-D}}
\newcommand{\trid}{\mbox{3-D}}
\newcommand{\der}{\ensuremath{\mathrm d}}
\newcommand{\clump}{\ensuremath{\mathrm{cl}}}
\newcommand{\mean}{\ensuremath{\mathrm{mean}}}

\newcommand{\zav}[1]{\left(#1\right)}
\newcommand{\hzav}[1]{\left[#1\right]}
\newcommand{\deriv}[2]{\ensuremath{\frac{{\mathrm d}#1}{{\mathrm d}#2}}}
\newcommand{\pderiv}[2]{\ensuremath{\frac{\partial #1}{\partial #2}}}

\newcommand{\Rstar}{\ensuremath{R_\ast}}
\newcommand{\Lstar}{\ensuremath{L_\ast}}
\newcommand{\Mstar}{\ensuremath{M_\ast}}

\newcommand{\Teff}{\ensuremath{T_\mathrm{eff}}}
\newcommand{\Kelvin}{\ensuremath{\mathrm{K}}}

\newcommand{\Lalpha}{\ensuremath{\mathrm{L}\alpha}}
\newcommand{\Halpha}{\ensuremath{\mathrm{H}\alpha}}
\newcommand{\Pdelta}{\ensuremath{\mathrm{P}\delta}}
\newcommand{\Pialpha}{\ensuremath{\mathrm{Pi}\alpha}}

\newcommand{\clfact}{\ensuremath{D}}
\newcommand{\volfilfact}{\ensuremath{f_\mathrm{vol}}}
\newcommand{\nelec}{\ensuremath{n_\mathrm{e}}}
\newcommand{\neleccl}{\ensuremath{({\nelec})_\clump}}
\newcommand{\nelecmean}{\ensuremath{({\nelec})_\mean}}
\newcommand{\nicl}{\ensuremath{{({n_i})_\clump}}}
\newcommand{\rhosm}{\ensuremath{\rho_\mathrm{sm}}}
\newcommand{\rhocl}{\ensuremath{\rho_\clump}}
\newcommand{\rhomean}{\ensuremath{\rho_\mean}}
\newcommand{\chism}{\ensuremath{\chi_\mathrm{sm}}}
\newcommand{\chicl}{\ensuremath{\chi_\clump}}
\newcommand{\chimean}{\ensuremath{\chi_\mean}}
\newcommand{\etacl}{\ensuremath{\eta_\clump}}
\newcommand{\etamean}{\ensuremath{\eta_\mean}}
\newcommand{\tauross}{\ensuremath{\tau_\mathrm{R}}}

\newcommand{\termls}[2]{\ensuremath{^{#1}\mathrm{#2}}}

\newcommand{\bfactor}{$b$\nobreakdash-factor}
\newcommand{\bfactors}{{\bfactor}s}

\newcommand{\Dlist}{\ensuremath{1,2,3,4,5,6,7,8}}
\newcommand{\Ddotlist}{\ensuremath{1,\dots,8}}

\defcitealias{ATA2}{II}
\defcitealias{ATA3}{Paper~III}
\defcitealias{ttbal}{KPP}

\begin{document} 

\title{Spherically symmetric model atmospheres using approximate lambda
operators}

\subtitle{V. Static inhomogeneous atmospheres of hot dwarf stars}

\author{Ji\v{r}\'{\i} Kub\'at\thanks{\href{https://orcid.org/0000-0003-4269-8278}
{https://orcid.org/0000-0003-4269-8278}}
\and
Brankica Kub\'atov\'a\thanks{\href{https://orcid.org/0000-0002-3773-2673}
{https://orcid.org/0000-0002-3773-2673}}
}

\institute{
        Astronomical Institute of the Czech Academy of Sciences,
        Fri\v{c}ova 298, CZ-251 65 Ond\v{r}ejov, Czech Republic
        \label{ondrejov}
}

\date{Received October 19, 2020; accepted June 7, 2021}

\abstract
{
Clumping is a common property of stellar winds and is being incorporated to a solution of the
radiative transfer equation coupled with kinetic equilibrium equations.
However, in static hot model atmospheres, clumping and its influence on the temperature and density
structures have not been considered and analysed at all to date.
This is in spite of the fact that clumping can influence the interpretation of resulting spectra, as
many inhomogeneities can appear there; for example, as a result of turbulent motions.
}
{
We aim to investigate the effects of clumping on atmospheric structure for the special case of a
static, spherically symmetric atmosphere assuming microclumping and a {\jednad} geometry.
}
{
Static, spherically symmetric, non-LTE (local thermodynamic equilibrium) model atmospheres were
calculated using the recent version of our code, which includes optically thin clumping.
The matter is assumed to consist of dense clumps and a void interclump medium.
Clumping is considered by means of clumping and volume filling factors,
assuming all clumps are optically thin.
Enhanced opacity and emissivity in clumps is multiplied by a volume filling factor to obtain their
mean values.
These mean values are used in the radiative transfer equation.
Equations of kinetic equilibrium and the thermal balance equation use clump values of densities.
Equations of hydrostatic and radiative equilibrium use mean values of densities.
}
{The atmospheric structure was calculated for selected stellar parameters.
Moderate differences were found in temperature structure.
However, clumping causes enhanced continuum radiation for the Lyman-line spectral region, while
radiation in other parts of the spectrum is lower, depending on the adopted model.
The atomic level departure coefficients are influenced by clumping as well.
}
{}


   \maketitle
%

\section{Introduction}

The calculation of {\jednad} model atmospheres has reached an extremely high level of
sophistication, as it is now possible to calculate NLTE (non-LTE; LTE means local thermodynamic
equilibrium) model atmospheres for different chemical compositions, with a lot of detail regarding
line and continuum formation, and, most importantly, with the inclusion of NLTE line blanketing
almost for all stellar types \citep[see][]{SA3}.
One of the most elaborate computer codes used for the calculation of NLTE, plane-parallel,
horizontally homogeneous model atmospheres is TLUSTY (see \citealt{Hubeny_2019} and references
therein).
However, all physical processes that lack the plane-parallel or spherical symmetry have to be
included in an approximate way.
As an example, convection is formed by a complex spectrum of turbulent motions.
However, it is usually treated using the mixing-length approximation.
Besides turbulent motions causing convection, it is the possible presence of simpler atmospheric
inhomogeneities, usually referred to as clumping, that are treated approximately as well.

Strictly speaking, as the {\jednad} model atmospheres are only homogeneous horizontally, they are in
fact inhomogeneous since the physical quantities vary with the depth coordinate ($z$ or $r$), but
this is not the meaning of an 'inhomogeneous atmosphere' in this paper.
Historically, the term `inhomogeneous stellar atmosphere' was also used to emphasise the horizontal
variability of quantities \citep[e.g.][]{Wilson_1968, Wilson_1969}. 
Clumping generally refers to the true inhomogeneity in all three dimensions, which is sometimes
referred to as the `stochastic medium' \citep[e.g.][]{Pomraning_1991, Gu_etal_1995}.

Clumping is a property of a stellar atmosphere, which is in principle {\trid}.
To describe its effects correctly, it has to be properly included using a {\trid} model atmosphere.
{\trid} stellar atmospheric modelling is being done for solar and cool star atmospheres.
Several powerful hydrodynamic codes exist to perform this task (e.g. \citealt{Stein_Nordlund_1998},
\citealt{Nordlund_Stein_2009}, \citealt{Freytag_etal_2012}, \citealt{Trampedach_etal_2013},
\citealt{Ludwig_Steffen_2016}, and \citealt{Freytag_etal_2019}).
For radiation-matter interaction, the assumption of the local thermodynamic equilibrium is being
used.
On top of these models, a NLTE {\trid} line-formation problem for particular ions is being
solved, mostly considering corresponding chemical elements as trace ones (e.g.
\citealt{Leenaarts_Carlsson_2009}, \citealt{Steffen_etal_2015}, \citealt{Bjorgen_Leenaarts_2017},
\citealt{Nordlander_etal_2017}, \citealt{Bergemann_etal_2019}, and references therein as recent
examples of calculations, or \citealt{Asplund_Lind_2010} for a review).

For the construction of hot star model atmospheres, the assumption of LTE is not acceptable at all
\cite[see discussions in][]{SA3}.
Hydrodynamics calculations in combination with NLTE are extremely demanding, and as such full {\trid}
treatment is still beyond the capabilities of contemporary computers, although recent
progress has been made \citep{Hennicker_etal_2018, Hennicker_etal_2020, Fisak_etal_2019}.
Therefore, {\trid} NLTE treatment of clumping in hot stars is only done in special cases; for
example, for radiative transfer in selected resonance lines \citep[e.g.][]{Sundqvist_etal_2010,
Sundqvist_etal_2011, clres1, modeling}.

It is widely accepted that clumping in line-driven stellar winds is caused by the
line-driven instability (often referred to as LDI).
Since its suggestion by \cite{Lucy_Solomon_1970}, subsequent analyses and modelling proved its
existence in {\jednad} hydrodynamic wind models (see \citealt{Owocki_etal_1988} and references
therein).
Hydrodynamical simulations of line-driven instabilities in stellar winds were recently extended to
{\dvad} \citep[see][and references therein]{Sundqvist_etal_2018, Owocki_Sundqvist_2018}.

However, the existence of a sub-photospheric convection in hot stars \citep{Cantiello_etal_2009,
Grassitelli_etal_2016} and other {\trid} effects in massive star envelopes
(\citealt{Jiang_etal_2015, Jiang_etal_2018} and references therein) offer another possibility for
the creation of clumps, which is not limited to a presence of a line-driven stellar wind and an
inherent instability.

The majority of wind modelling codes consider clumping in a {\jednad} approximation, which means that
severe simplifications have to be employed.
The assumption of a {\jednad} spherically symmetric flow is employed in the NLTE wind modelling
codes, which solve the NLTE equations (i.e. radiative transfer and kinetic equilibrium equations)
and optionally an equation for temperature determination (radiative equilibrium or thermal balance
equations) for a given density and velocity structure.
To our knowledge, three such codes exist, namely CMFGEN \citep[e.g.][]{Hillier_Miller_1998},
PoWR \citep[e.g.][]{Hamann_Grafener_2004}, and FASTWIND \citep[e.g.][]{Santolaya-Rey_etal_1997,
Puls_etal_2005}.
These codes consider clumping as a simple correction of the existing smooth hydrodynamic structure
using the clumping factor or its inverse (the volume filling factor), and solve the NLTE line
formation problem in such a modified atmosphere without changing the mean atmospheric structure.
Many clumped wind models have been calculated in such way.

The actual value of the adopted clumping factor is a matter of discussion.
Sometimes it is necessary to use very high clumping factors \citep[e.g. $\sim50$ for WR
stars,][]{Sander_Vink_2020}, but this necessity is probably caused by the fact that these clumping
factors refer to microclumping.
If macroclumping (clumps are allowed to be optically thick) is taken into account, much smaller
values are sufficient \citep{wr136}.

Attempts to include the influence of clumping on the mean structure of a {\jednad} stellar
atmosphere (mean density, velocity) are not so frequent.
\cite{lianne} studied the effect of clumping on mass-loss rate predictions.
Wind models with optically thin clumps both in lines and continuum (microclumping) were used in a
calculation presented by (\citealt{irchuch}, see also references therein).
A method for inclusion of clumping into rate equations for {\jednad} NLTE wind models was developed
by \cite{Sundqvist_etal_2014} and used by \cite{Sundqvist_Puls_2018}.

In the past, we developed a static {\jednad} model atmosphere code \citep[and references
therein]{ATAsum} capable of building the structure of either spherically symmetric or plane-parallel
NLTE model atmospheres.
Given the evidence that clumping in hot star atmospheres can be caused by processes not connected
with the existence of stellar winds (e.g. sub-photospheric convection), we decided to include an
approximate description of atmospheric inhomogeneities to {\jednad}
static model atmosphere construction, which has not been done before,
and to study the consequences of such an approach.

\section{Description of clumping}
\label{section_clump}

Stellar atmospheres are very often modelled using a simplified picture.
The most commonly used model is the atmosphere consisting of horizontally homogeneous layers, which
may be parallel layers for plane-parallel atmospheres or concentric shells for spherically symmetric
atmospheres.
This offers a possibility to use the advantage of a {\jednad} geometry description.
However, as direct observations of selected stars show surface structures, the assumption of
horizontally homogeneous atmospheres is questioned.
It is highly probable that no stars have strictly horizontally homogeneous atmospheres and their
inhomogeneity is significantly more complicated.
Some parts of the atmosphere can violate the assumption of horizontal homogeneity having either
higher or lower density than the values predicted by a horizontally homogeneous model, with a
variable size of these regions.
Since the details of stellar atmosphere inhomogeneities and process of their formation are vastly
unknown and their description involving the full spectrum of possible shapes is far complicated, a
simplified description of inhomogeneity is necessary.
In {\jednad} model atmospheres this is conveniently done using clumping or volume filling factors.
We describe the {\jednad} model of clumping applied in this paper in a more detail.

\subsection{{\jednad} treatment of clumping}

The atmosphere is assumed to consist of clumps.
To simplify the problem, the space between clumps is assumed to be void.
Clumps are assumed to be smaller than the photon mean free path, and this assumption is often
referred to as optically thin clumping \citep{Hamann_Koesterke_1998} or microclumping
\citep{Oskinova_etal_2007}.
Clumping is usually characterised by a clumping factor (we denote this quantity as $\clfact$ in
this paper), a ratio of a mean-square deviation of the density and the mean density squared
\citep[see][Eq.~25] {Chandrasekhar_Munch_1952}, which, for the case of a {\jednad,} spherically
symmetric stellar atmosphere, can be written introducing its radius dependence as
\begin{equation}\label{clchand}
\clfact(r) = \frac{\langle\rho^2(r)\rangle}{\langle\rho(r)\rangle^2}
,\end{equation}
or using $z$ instead of $r$ as the independent variable for the case of the plane-parallel
atmosphere.
Introducing the volume filling factor $\volfilfact(r)$, which describes the fraction of volume
filled by clumps at a radius $r$, we can relate the density in clumps $\rhocl(r)$ to the mean
density of the atmosphere $\rhomean(r)$ as
\begin{equation}\label{rhomeandef}
\rhomean(r) = \volfilfact(r) \rhocl(r).
\end{equation}
Inserting this equation into \eqref{clchand}, we obtain \citep[see also][]{Sundqvist_Puls_2018}
\begin{equation}\label{clfactdef}
\clfact(r) = \frac{\rhocl(r)}{\rhomean(r)}.
\end{equation}
In other words, $\rhomean$ is the density as it would be in a medium without clumps (in a
horizontally homogeneous atmosphere).
This quantity is also being referred to as the smooth atmosphere (wind) density ($\rhosm=\rhomean$).
The equation \eqref{clfactdef} is valid also for all variables describing number densities.
For example, the ratio of the electron number density inside clumps $\neleccl$ and the mean electron
number density {\nelecmean} is also $D(r)$.

It has to be noted that the {\jednad} description of clumps may be more detailed.
If the assumption of optically thin clumping is violated, then more parameters are necessary to
describe properties of a clumped medium.
The quantity porosity length $h$ describing the photon mean free path between clumps (which may be
optically thick) has to be introduced to the clump description \citep[see][]{Feldmeier_etal_2003,
Owocki_etal_2004, Sundqvist_Puls_2018}.
However, the porosity length $h$ is an additional free parameter, and we aim to test basic
clumping effects in optically thin ($\tauross \lesssim 2/3$) parts of static {\jednad} atmospheres.
To avoid using too many free parameters for this purpose, we decided to postpone the more detailed
analysis using the porosity length to future, and thus we keep the simpler assumption of optically
thin clumping in this paper.

In the following sections, references to equations from \citet[hereafter
Paper~\citetalias{ATA2}]{ATA2} are made.
These equations are denoted as \citetalias{ATA2}.x.

\subsection{Opacity, emissivity, and scattering}

Opacity, emissivity, and scattering coefficients are considered per volume in this paper.
Inside clumps, they are evaluated using actual mass and number densities there.
The total opacity in clumps {\chicl} is given as a sum of atomic opacities (bound-bound, bound-free,
and free-free) and the scattering opacity,
\begin{equation}\label{totclop}
\chicl = \zav{\chicl^\mathrm{bb} + \chicl^\mathrm{bf} + \chicl^\mathrm{ff}} + \chicl^\mathrm{sc}
= \chicl^\mathrm{at} + \chicl^\mathrm{sc},
\end{equation}
whereas the total emissivity in clump {\etacl} consists of bound-bound, bound-free, and
free-free emissivities, and the following scattering emissivity:
\begin{equation}\label{totclem}
\etacl = \zav{\etacl^\mathrm{bb} + \etacl^\mathrm{bf} + \etacl^\mathrm{ff}} 
+ \etacl^\mathrm{sc} =
\etacl^\mathrm{at} + \etacl^\mathrm{sc}.
\end{equation}
Naturally, the opacity, emissivity, and scattering coefficients in the void interclump medium are
zero.
To obtain mean values of the coefficients, which are the means over both clumps and void
interclump medium, we simply have to multiply the opacity in the clumps by the volume filling
factor, namely
\begin{equation}\label{meanclumpfact}
\begin{aligned}
\chimean & = \volfilfact \chicl,
\\
\etamean & = \volfilfact \etacl.
\end{aligned}
\end{equation}
According to Eq.~\eqref{clfactdef}, the densities in clumps are $\clfact$ times larger than the
densities in an atmosphere without clumps ($\rhocl = \clfact \rhosm$).
As a result, in clumps, all processes whose opacity depends on the first power of density, like all
line absorptions and emissions, photoionisation, and scattering on free electrons, have $\clfact$
times larger values of opacity than in the atmosphere without clumps (e.g. for bound-bound
transitions $\chicl^\mathrm{bb} = \clfact \chism^\mathrm{bb}$).
The mean opacity of the medium is obtained as a volume weighted sum of opacities in clumps and in an
interclump medium.
Consequently, for a void interclump medium the expression for bound-bound opacities simplifies to
$\chimean^\mathrm{bb} = \volfilfact \chicl^\mathrm{bb} = \volfilfact \clfact \chism^\mathrm{bb} =
\chism^\mathrm{bb}$), and similarly for the bound-bound emissivity $\etamean^\mathrm{bb}$,
bound-free opacity $\chimean^\mathrm{bf}$, and electron scattering $\chicl^\mathrm{sc}$.
Processes, whose opacity depends on a product of two densities, like
free-free transitions or radiative recombination (depend on a product of
electron density and ion density), have in clumps $\clfact^2$  larger values of
opacity than in the atmosphere without clumps
(e.g. for free-free opacity $\chicl^\mathrm{ff} = \clfact^2 \chism^\mathrm{ff}$).
Consequently, the corresponding mean opacity $\chimean^\mathrm{ff} = \volfilfact \chicl^\mathrm{ff}
= \volfilfact \clfact^2 \chism^\mathrm{ff} = \clfact \chism^\mathrm{ff}$) and similarly for
$\etamean^\mathrm{ff}$ and $\etamean^\mathrm{fb}$ \cite[see][Eq.~10]{statcl}.
This also affects other quantities appearing in the radiative transfer equation, such as the optical
depth and the source function, whose mean values have more complicated density dependence because
they depend on a combination of processes with different clumping dependences.

\subsection{Radiative transfer equation}

To include the atmospheric inhomogeneities in the {\jednad} case, the planar (\citetalias{ATA2}.1)
and the spherically symmetric (\citetalias{ATA2}.2) radiative transfer equations have to be modified
by using mean values of opacity and emissivity \eqref{meanclumpfact}, namely
\begin{subequations}
\begin{equation}\label{rtepp}
\mu \deriv{I\zav{\nu,\mu}}{z} = - \chimean(\nu) I\zav{\nu,\mu} + \etamean(\nu)
\end{equation}
for the planar case, and
\begin{equation}\label{rterf}
\mu \pderiv{I\zav{\nu,\mu}}{r} + {1 - \mu^2 \over r} \pderiv{I\zav{\nu,\mu}}{\mu}
= - \chimean(\nu) I\zav{\nu,\mu} + \etamean(\nu)
\end{equation}
\end{subequations}
for the spherical geometry.
Associated moment equations for the mean intensity read
\begin{subequations}
\begin{equation}\label{rtempp}
\deriv{^2\hzav{f(\nu) J(\nu)}}{\hzav{\tau(\nu)}^2} = J(\nu) - S(\nu)
\end{equation}
for the planar geometry, and
\begin{equation}\label{rtemsph}
\deriv{^2\hzav{f(\nu) q(\nu) J(\nu)}}{\hzav{X(\nu)}^2} = \frac{r^4}{q(\nu)} \hzav{J(\nu) - S(\nu)}
\end{equation}
\end{subequations}
for the spherical geometry; $q(\nu)$ is the sphericity factor \citep{Auer_1971} defined by equation
(\citetalias{ATA2}.3),
and $f(\nu)$ is the variable Eddington factor.
Here, the plane-parallel optical depth is
\begin{subequations}
\begin{equation}\label{taunu}
\der \tau(\nu) = - \chimean(\nu) \der r,
\end{equation}
the optical-depth-like variable $X(\nu)$ used in the spherically symmetric case
\citep[different from the one used by][]{Mihalas_Hummer_1974} is
\begin{equation}\label{Xnu}
\der X(\nu) = - \frac{q(\nu) \chimean(\nu)}{r^2} \der r,
\end{equation}
\end{subequations}
and the source function is
\begin{equation}
S(\nu) = \frac{\etamean(\nu)}{\chimean(\nu)}.
\end{equation}

\subsection{Equations of kinetic (statistical) equilibrium}

The equations of kinetic (statistical) equilibrium are evaluated using actual density $\rhocl$ in
clumps, which means also using the actual electron density $\neleccl$.
The equations are then solved for $\nicl$, the occupation numbers for matter inside clumps.
Values of both collisional and radiative rates are calculated using proper values of densities.

\subsection{Structural equations}

The structural equations determining the temperature and density structure of the atmosphere are
treated using mean values of densities.
This is important for the equation of hydrostatic equilibrium and the differential form of the
equation of radiative equilibrium.

The basic independent variable of the model is the column mass depth, which is defined as
$\der m = - \rho \der z$ for the plane-parallel atmosphere and
$\der m = - \rho \zav{\Rstar^2 / r^2} \der r$ for the spherically symmetric
atmosphere ($\Rstar$ is the stellar radius, see \citetalias{ATA2}.9).
For the case of a clumped model atmosphere, we define the column mass depth as
\begin{subequations}\label{dmdef}
\begin{equation}
\der m = - \rhomean \der z
\end{equation}
for the plane parallel case, and
\begin{equation}
\der m = - \rhomean \zav{\frac{\Rstar^2}{r^2}} \der r
\end{equation}
\end{subequations}
for the spherically symmetric case ({\rhomean} was introduced in Eq.~\ref{rhomeandef}).

\subsubsection{Radiative equilibrium}

The integral form of the equation of radiative equilibrium expresses the balance between absorbed
and emitted radiation energy.
Consequently, it remains the same as in the case of homogeneous atmospheres, namely it has the form
(\citetalias{ATA2}.12), and the opacity and emissivity values in clumps are used:
\begin{equation}\label{ere1}
\int_0^\infty \hzav{\chicl^\mathrm{at}(\nu) J_\nu - \etacl^\mathrm{at}(\nu)} \der \nu = 0,
\end{equation}
where $J_\nu$ is the mean radiation intensity,
$\nu$ stands for frequency,
and $\chicl^\mathrm{at}$ and $\etacl^\mathrm{at}$ were introduced in Eq.\,\eqref{totclop}.
As the interclump medium is void in our case, the equation of radiative equilibrium has no physical
sense there.
Taking into account Eq.~\eqref{meanclumpfact}, mean opacity and emissivity
values can be used in Eq.~\eqref{ere1} as well.

The differential form of the equation of radiative equilibrium (\citetalias{ATA2}.19) describes the
conservation of the total radiative flux, which flows through both clumps and the interclump medium.
Consequently, mean quantities have to be used.
We consider its modified form:
\begin{equation}\label{eredif3}
H_0 = \int_0^\infty \frac{\rhomean}{q(\nu) {\chimean}(\nu)}
\deriv{\hzav{q(\nu) f(\nu) J(\nu)}}{m} \der\nu
,\end{equation}
where
${\chimean}(\nu)$ is the mean opacity (absorption + scattering),
$q(\nu)$ is the sphericity factor (see Eq.\,\ref{rtemsph})
for the spherically symmetric case (for the plane-parallel case we may set $q(\nu) = 1$),
$f(\nu)$ is the variable Eddington factor, and
$H_0 = \sigma \Teff^4 / (4 \pi )$ for the plane-parallel case and
$H_0 = \Lstar / ( 4 \pi \Rstar )^2$ for the spherically symmetric case.
$\Lstar$ is the stellar luminosity, and $\Teff$ is the stellar effective temperature.

\subsubsection{Thermal balance}

Equations of thermal balance of electrons \citep[see][hereafter \citetalias{ttbal}]{ttbal} can
replace the integral form of the radiative equilibrium in the outermost layers of the stellar
atmospheres, where the bound-bound radiative rates are so strong that they numerically saturate the
radiative equilibrium equation \eqref{ere1}, and their difference is subject to a large numerical
error.

Thermal balance equations describe the local balance of thermal energy.
Consequently, all number densities in the equations in \citetalias{ttbal} have to be replaced by
their clump values\footnote{We note that in \cite{statcl} the mean number density values were
incorrectly used in thermal balance equations.}.
Free-free heating and cooling are expressed using equations (\citetalias{ttbal}.3) and
(\citetalias{ttbal}.4), respectively, where $\nelec \rightarrow \neleccl$ and $N_j \rightarrow
(N_j)_\clump$.
Similarly, for bound-free heating and cooling equations (\citetalias{ttbal}.5) and
(\citetalias{ttbal}.6), respectively, are used with $n_l^\ast \rightarrow (n_l^\ast)_\clump$.
For collisional heating and cooling, equations (\citetalias{ttbal}.11) and (\citetalias{ttbal}.12)
are used with the above-mentioned substitutions.

\subsubsection{Hydrostatic equilibrium}

Using \eqref{dmdef}, the equation of hydrostatic equilibrium can be written in the same form as
(\citetalias{ATA2}.10)
\begin{equation}\label{ehe1}
\deriv{p_g}{m} = \frac{G \Mstar}{\Rstar^2}
- \frac{4 \pi}{c} \int_0^\infty \frac{1}{q(\nu)}
\deriv{\hzav{q(\nu) f(\nu) J(\nu)}}{m} \der\nu,
\end{equation}
where $p_g$ is the gas pressure,
{\Mstar} is the stellar mass, 
and $c$ is the light speed.
This equation is accompanied by the upper boundary condition (\citetalias{ATA2}.11),
\begin{equation}\label{ehe1up}
\frac{p_1}{m_1} = \frac{G \Mstar}{\Rstar^2}
- \frac{4 \pi}{c} \zav{\frac{r_1}{\Rstar}}^2 \int_0^\infty \frac{\chi_1(\nu)}{\rho_1}
\hzav{g_1(\nu) J_1(\nu) - H^-(\nu)} \der\nu,
\end{equation}
where the index $1$ denotes quantities at the uppermost depth point of the model
(we set $r_1=\Rstar$ for the plane-parallel case),
$g_1(\nu)= \int_0^1\mu j_1(\nu,\mu) \der \mu / J_1(\nu)$
($j(\nu,\mu)$ is the mean intensity-like \cite{Schuster_1905} variable),
and $H^-(\nu)$ is the incident flux.
The opacity $\chi$ and density $\rho$ are both either clump or mean values, since the clumping
factors are cancelled in the fraction.

\subsubsection{Clumping at great depths}

On average, clumping increases the opacity thanks to the $D^2$ dependence of the free-free
absorption coefficient.
Consequently, for a fixed column-mass-depth scale $m$ given by Equation \eqref{dmdef}, the optical
depth of a clumped medium increases for all frequencies.
Depending on the clumping factor $D$, the depth of radiation formation shifts upwards, which affects
the temperature structure.
At great depths, clumping causes an increase of the Rosseland optical depth $\tauross$, temperature
follows the gray dependence \citep[e.g.][Eq.~17.67]{SA3}, and it results in some heating there.

There is, of course, a question of whether clumping in our approximation of void interclump medium
and optically thin clumping has any physical meaning at significant optical depths ($\tau \gg 1$).
There, the diffusion approximation for radiation is valid, but assuming horizontally homogeneous
medium.
A void interclump medium means that radiation can propagate without any interaction with matter
through these interclump voids, which contradicts radiation diffusion.
To avoid these unrealistic effects on the deepest (and optically thick) parts of the atmosphere, we
decided to assume no clumping there by setting $D=1$ for $\tauross \gtrsim 2/3$ with gradual
transition using linear interpolation in $\log\tauross$ to actual $D$ at $\tauross \sim 0.1$.

\subsection{Implementation in the code}

Our method for calculation of {\jednad} static NLTE model atmospheres and the associated code
\citep{ATA1, ATA2, ATA3, ATA4, ATAsum} use the accelerated lambda iteration technique in combination
with the complete linearisation (multi-dimensional Newton-Raphson) method.
The code had to be updated to include clumping as described in this section.
Basic changes are described below.

The basic independent variable $m$ (column mass depth) is defined using mean densities after
\eqref{dmdef}.
After the $m$-scale is set, it is fixed during all calculations.
For densities, both values inside clumps ($\rhocl$, $\neleccl$) and mean values ($\rhomean$,
$\nelecmean$) are stored for each depth point.
The relation between $\rhocl$ and  $\rhomean$ is given by \eqref{clfactdef}, similarly to the
electron number density.
Alternatively, only one value may be stored and the other recalculated when necessary, but we
decided to store both values.

Opacities are first evaluated using actual densities in clumps ($\rhocl$, $\neleccl$,
...).
Then, the opacities are divided by the clumping factor (Eq.~\ref{meanclumpfact}) to obtain their mean
values.
Next, the Rosseland and flux mean opacities are calculated.
Mean opacities and emissivities \eqref{meanclumpfact} are used in the radiative transfer equation to
take into account transfer of radiation through a mean medium consisting of both clumps and
an interclump medium.  

Kinetic equilibrium equations are formulated using number densities (occupation numbers) in clumps.
Electron number densities that enter the equations for radiative and collisional rates are clump
electron number densities $\neleccl$.

The temperature is that of matter in clumps.
Two of the temperature determination methods, namely the integral form of radiative equilibrium and
the thermal balance method, describe local processes.
Opacities calculated from densities in clumps ($\rhocl$, $\neleccl$) were used.
The differential form of radiative equilibrium uses mean values of densities ($\rhomean$), since
this equation describes conservation of the radiation flux.
Mean densities $\rhomean$ are also used in the equation of hydrostatic equilibrium.

\section{Model calculations}

\newcommand{\sdpetpet}{{\tt 55-4}}
\newcommand{\Osest}{{\tt O6}}
\newcommand{\Oosm}{{\tt O8}}
\newcommand{\Bnula}{{\tt B0}}
\newcommand{\Bnulapet}{{\tt B0.5}}
\newcommand{\Bdva}{{\tt B2}}
\newcommand{\Bctyri}{{\tt B4}}
\newcommand{\Bsest}{{\tt B6}}

\begin{table*}
\caption{Basic input global parameters of calculated model atmospheres.}
\label{parametry_modelu}
\centering
\begin{tabular}{lcccccc}
\hline
model & $L[\Lstar]$ & $R[\Rstar]$ & $M[\Mstar]$ & $\Teff [\Kelvin]$ & $\log g [\mathrm{cm} /
\mathrm{s}^2]$ & ions
\\
\hline
\sdpetpet & $1.321 \cdot 10^{3}$ & $0.40$ & $0.500$ & $54\,999$ & $4.933$ &
 \ion{H}{i}, \ion{He}{i}, \ion{He}{ii} \\
\Osest    & $2.643 \cdot 10^{5}$ & $9.85$ & $31.65$ & $41\,690$ & $3.951$ &
 \ion{H}{i}, \ion{He}{i}, \ion{He}{ii} \\
\Oosm     & $8.139 \cdot 10^{4}$ & $7.51$ & $21.66$ & $35\,560$ & $4.022$ &
 \ion{H}{i}, \ion{He}{i}, \ion{He}{ii} \\
\Bnula    & $2.411 \cdot 10^{4}$ & $5.80$ & $14.57$ & $29\,850$ & $4.074$ &
 \ion{H}{i}, \ion{He}{i}, \ion{He}{ii} \\
\Bdva     & $4.722 \cdot 10^{3}$ & $4.28$ & $8.620$ & $23\,120$ & $4.111$ &
 \ion{H}{i}, \ion{He}{i}, \ion{He}{ii} \\
\Bctyri   & $8.351 \cdot 10^{2}$ & $3.26$ & $5.120$ & $17\,180$ & $4.121$ &
 \ion{H}{i}, \ion{He}{i} \\
\Bsest    & $2.810 \cdot 10^{2}$ & $2.81$ & $3.800$ & $14\,090$ & $4.120$ &
 \ion{H}{i}, \ion{He}{i} \\
\hline
\end{tabular}
\end{table*}

Using our modified code, we calculated several test model sets of static atmospheres.
Model atmospheres consist only of hydrogen and helium.
Atomic data and model atoms were the same as thosed we used previously \citep{mp3}.
The helium abundance in all models was $Y_\mathrm{He} = n_\mathrm{He} / n_\mathrm{H} = 0.1$,
$n_\mathrm{He}$ and $n_\mathrm{H}$ are mean number densities of helium and hydrogen, respectively.
Basic global stellar parameters (luminosity $\Lstar$, radius $\Rstar$, mass $\Mstar$, effective
temperature $\Teff$, surface gravity $g$) of calculated models are listed in
Table~\ref{parametry_modelu}.
These parameters were chosen to cover B- and O-type main-sequence stars \citep[{\Osest}, {\Oosm},
{\Bnula}, {\Bdva}, {\Bctyri}, and {\Bsest} parameters after][]{Harmanec_1988}, supplemented by one
hot subdwarf model \citep[{\sdpetpet}, parameters as in][]{snehurka}.
For each set of parameters, models for clumping factors $\clfact = \Dlist$ were
calculated.

We calculated full static spherically symmetric NLTE model atmospheres, meaning solutions for
electron density $\nelec$, temperature $T$, radius $r$, and number densities for all atomic levels
$n_i$ ($i=1, \dots, L$, where $L$ is the total number of atomic levels considered).
In the model calculation output, the latter are expressed as products of LTE number densities
$n_i^*$ (for actual values of temperature and mass density) and departure coefficients ($b$-factors
$b_i = n_i / n_i^*$).

The depth scale of our models was from $m\sim 10^3$ to $m\sim 10^{-8}$, and in special cases to $m
\sim 10^{-9}$ (in CGS units).
We chose the outer boundary at low $m$ (consequently low Rosseland optical depth $\tauross$, which
is of the same order of magnitude) to ensure that all explicitly considered lines  become optically
thin within the model.
All models were iterated until the relative change for all iterated variables was lower than
$10^{-4}$.

We encountered convergence problems for high clumping factors for models {\Bdva} (see
Table~\ref{parametry_modelu}, to overcome them we set the resonance transitions of helium ions
(\ion{He}{i} $1s\,\termls{1}{S} \leftrightarrow 2p\termls{1}{P}$, \ion{He}{ii} $n=1\leftrightarrow
n=2$) to a detailed radiative balance.
To obtain a consistent set of models, we used this detailed radiative balance approximation for all
clumping factors for the {\Bdva} models.

\section{Results}

\subsection{Temperature structure}

\begin{figure*}
\resizebox{0.5\hsize}{!}{\includegraphics{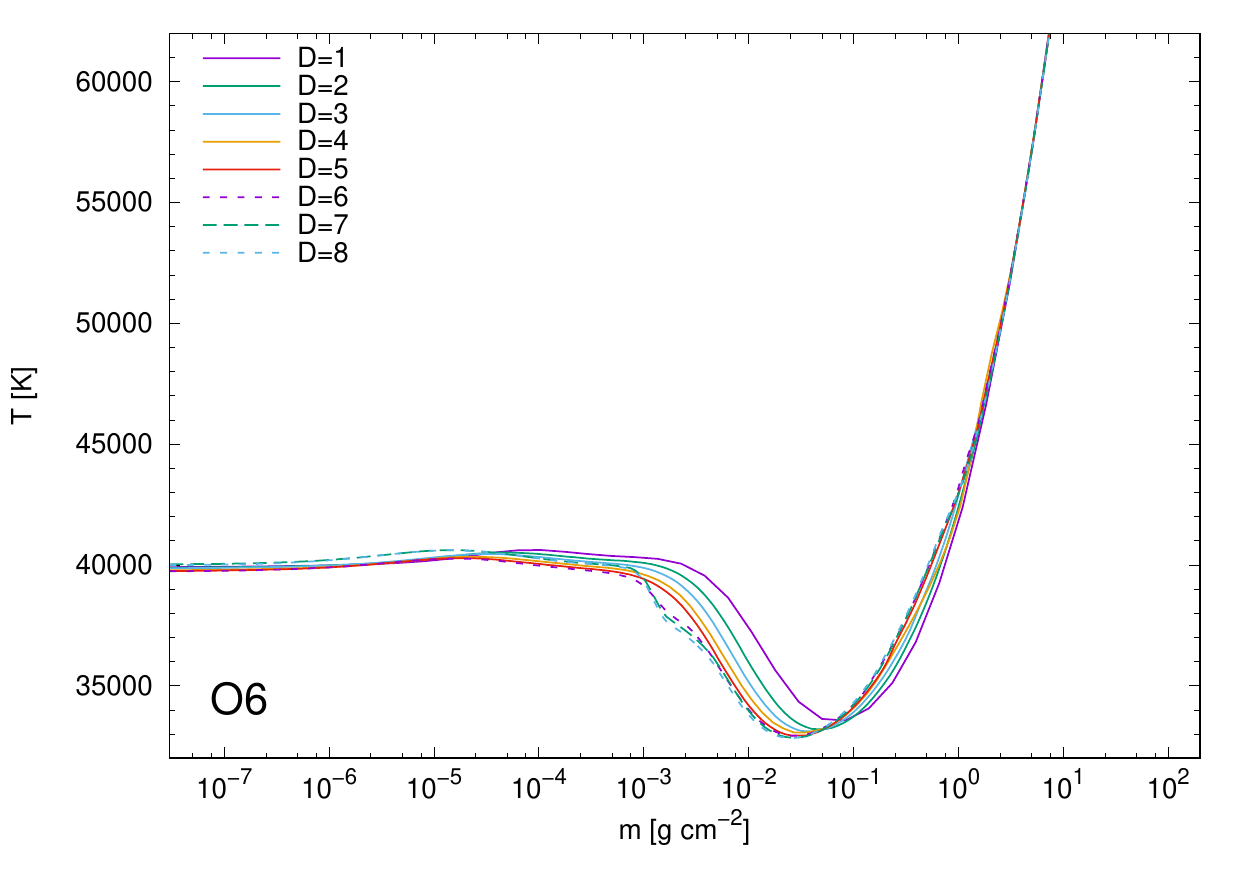}}
\resizebox{0.5\hsize}{!}{\includegraphics{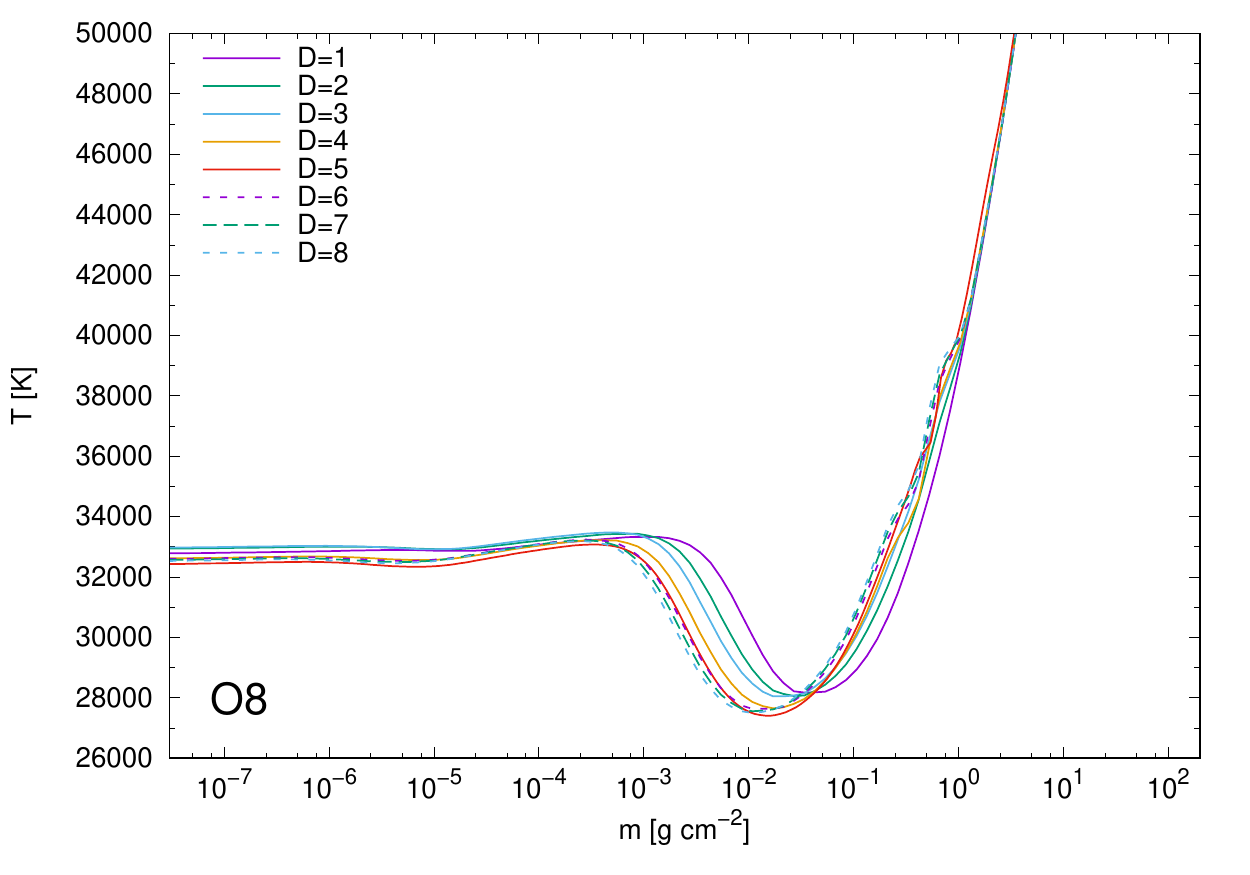}}
\resizebox{0.5\hsize}{!}{\includegraphics{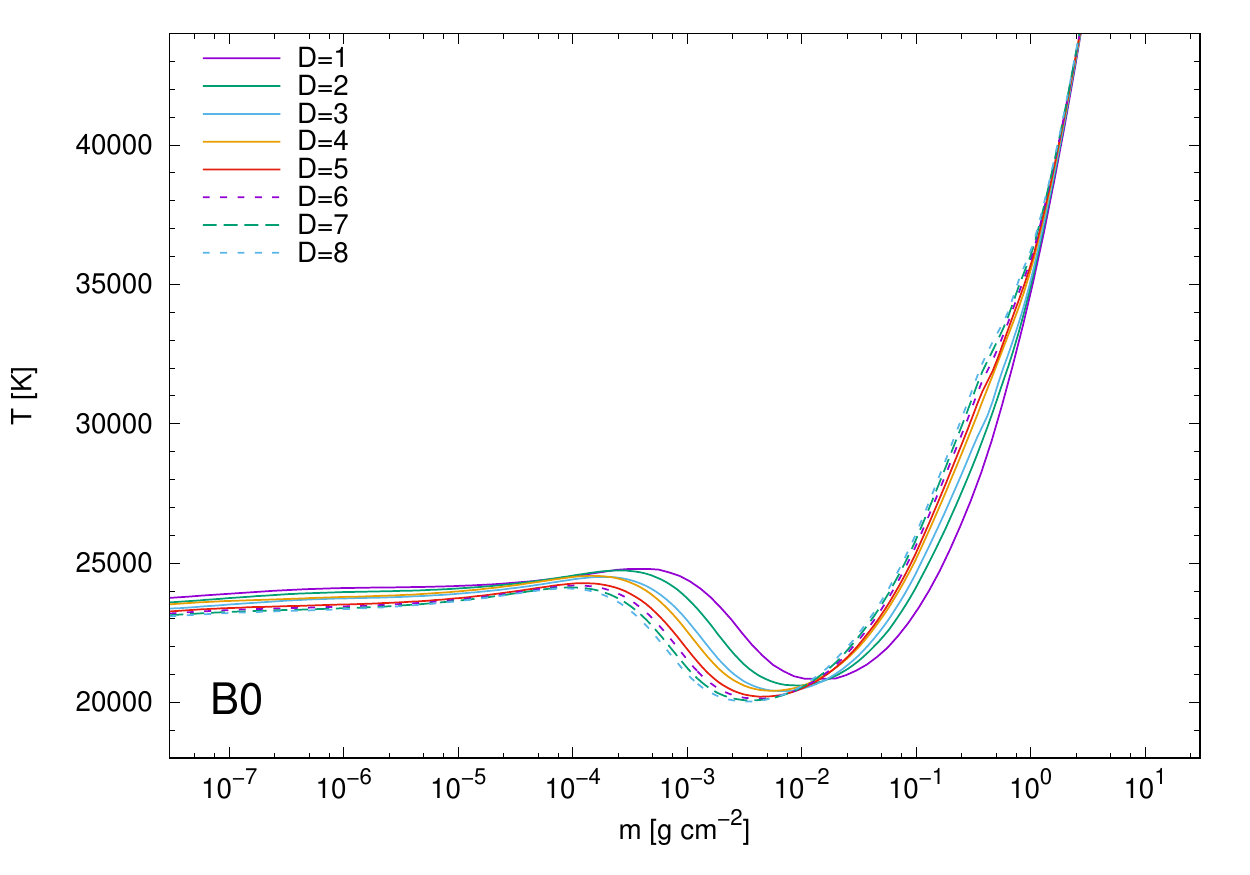}}
\resizebox{0.5\hsize}{!}{\includegraphics{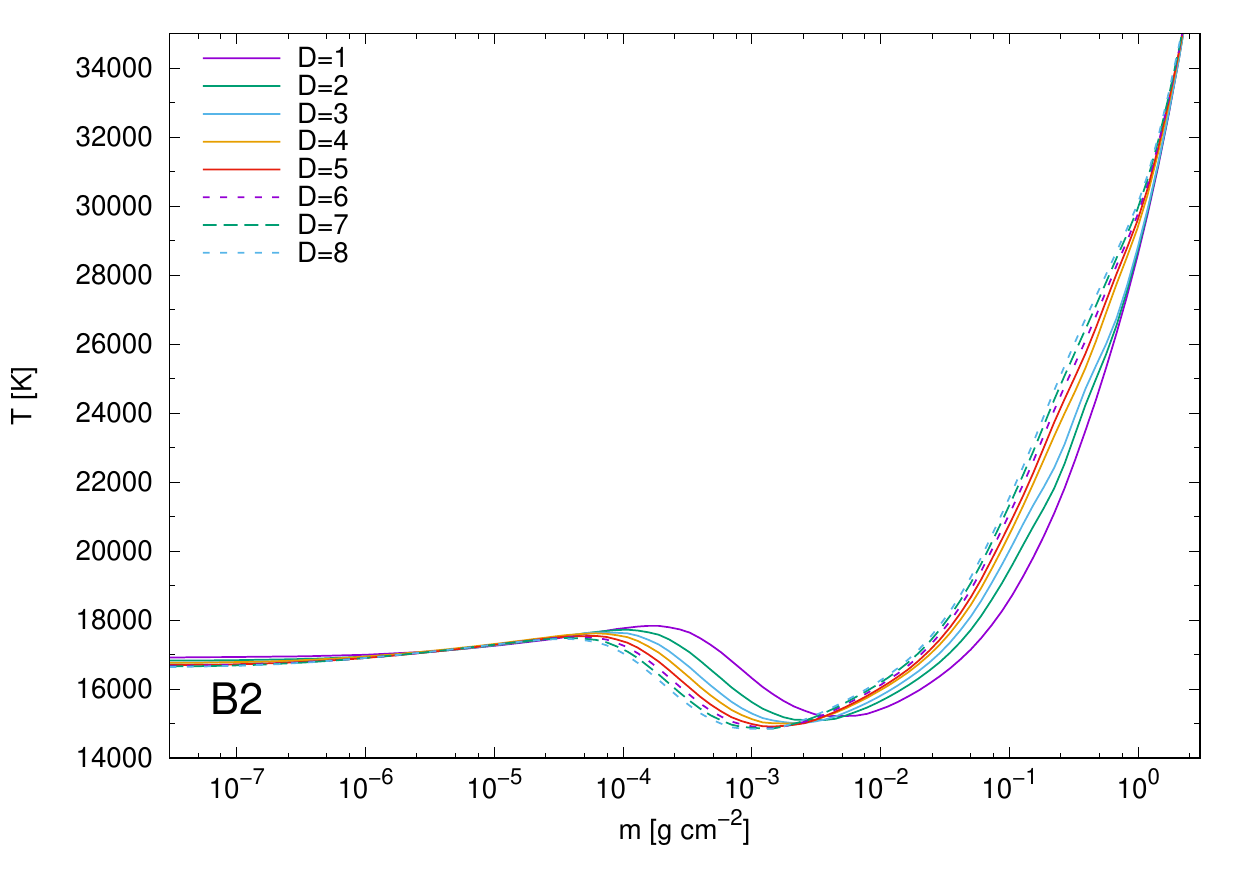}}
\resizebox{0.5\hsize}{!}{\includegraphics{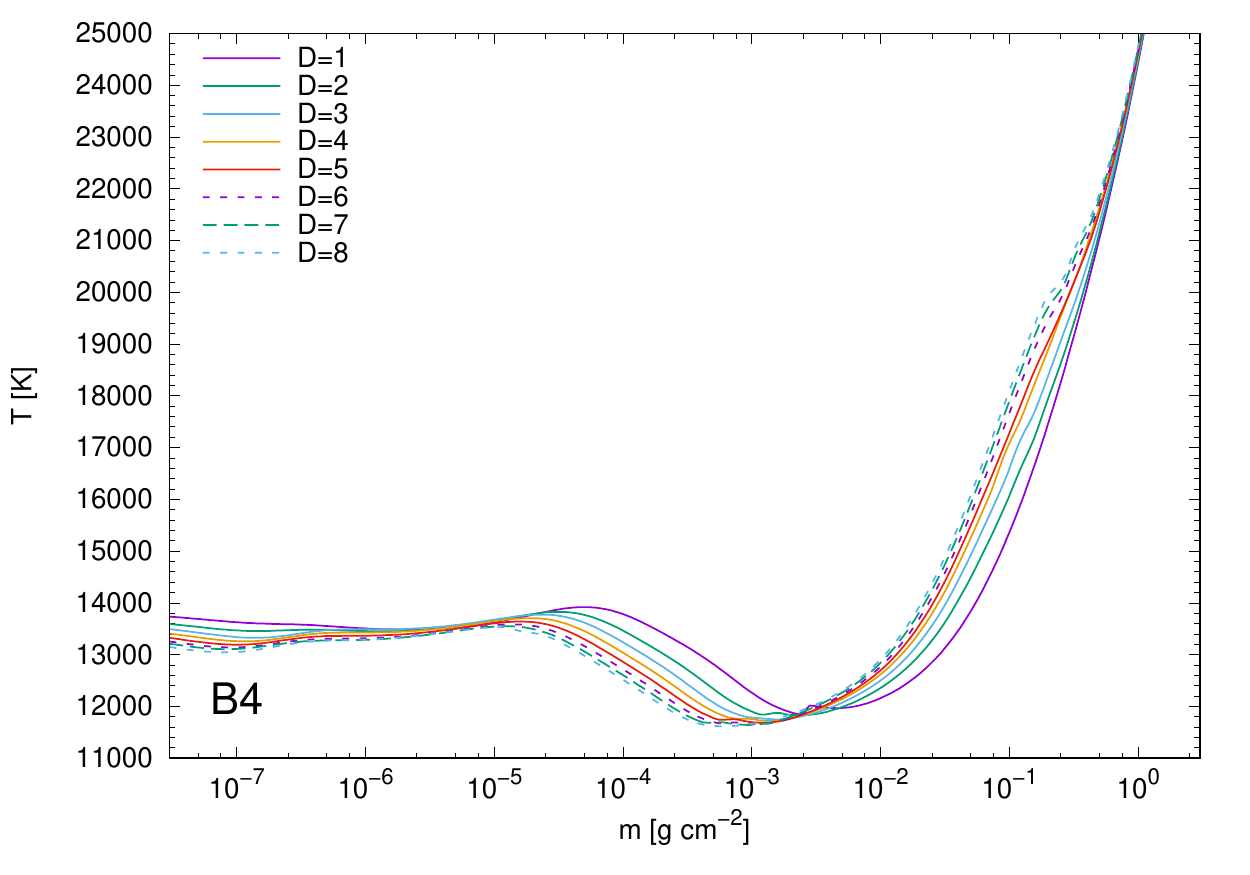}}
\resizebox{0.5\hsize}{!}{\includegraphics{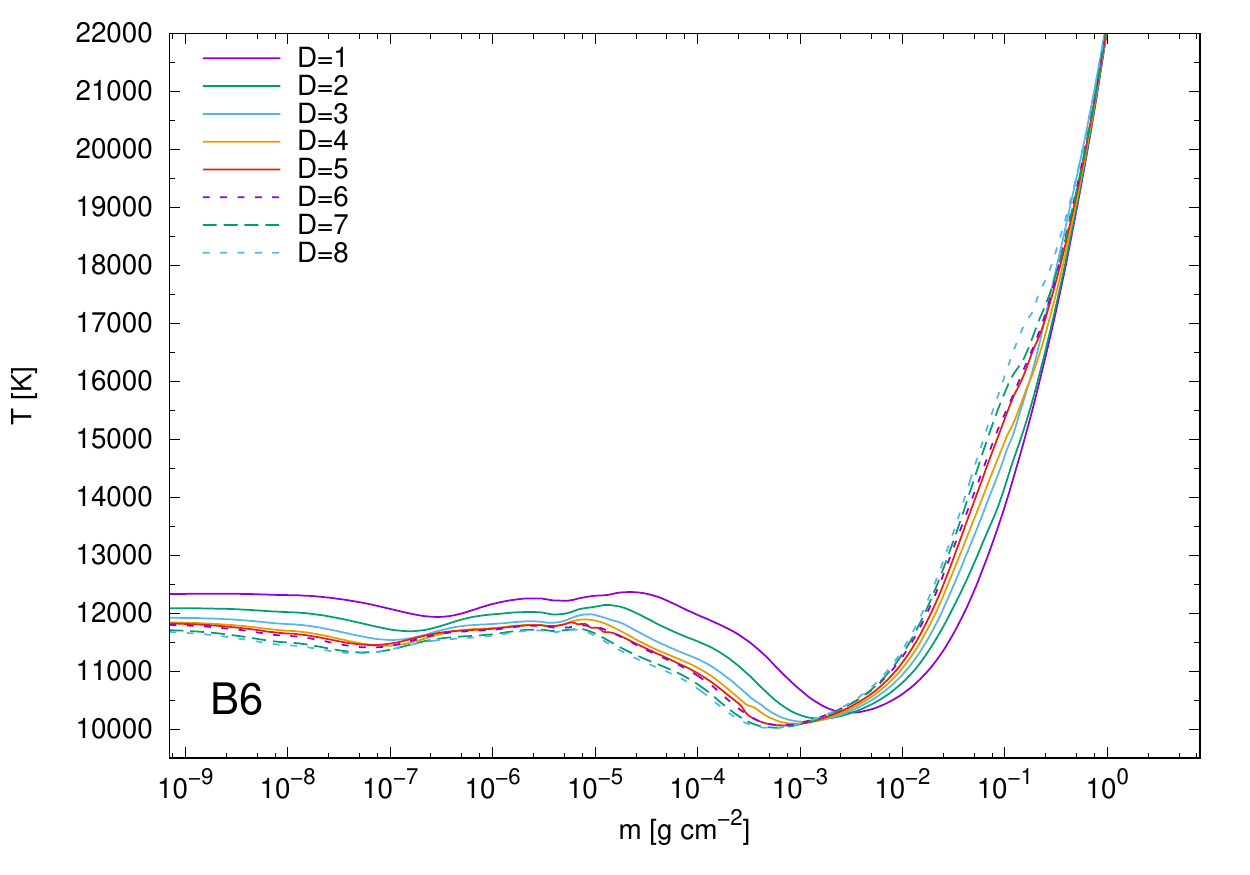}}
\caption{
Temperature structure for models
{\Osest} (upper left panel),
{\Oosm} (upper right panel),
{\Bnula} (middle left panel),
{\Bdva} (middle right panel),
{\Bctyri} (lower left panel),
and
{\Bsest} (lower right panel)
for all values of the clumping factor ($\clfact = \Ddotlist$).}
\label{O6O8B05B2B4B6teplota}
\end{figure*}

\begin{figure}
\resizebox{\hsize}{!}{\includegraphics{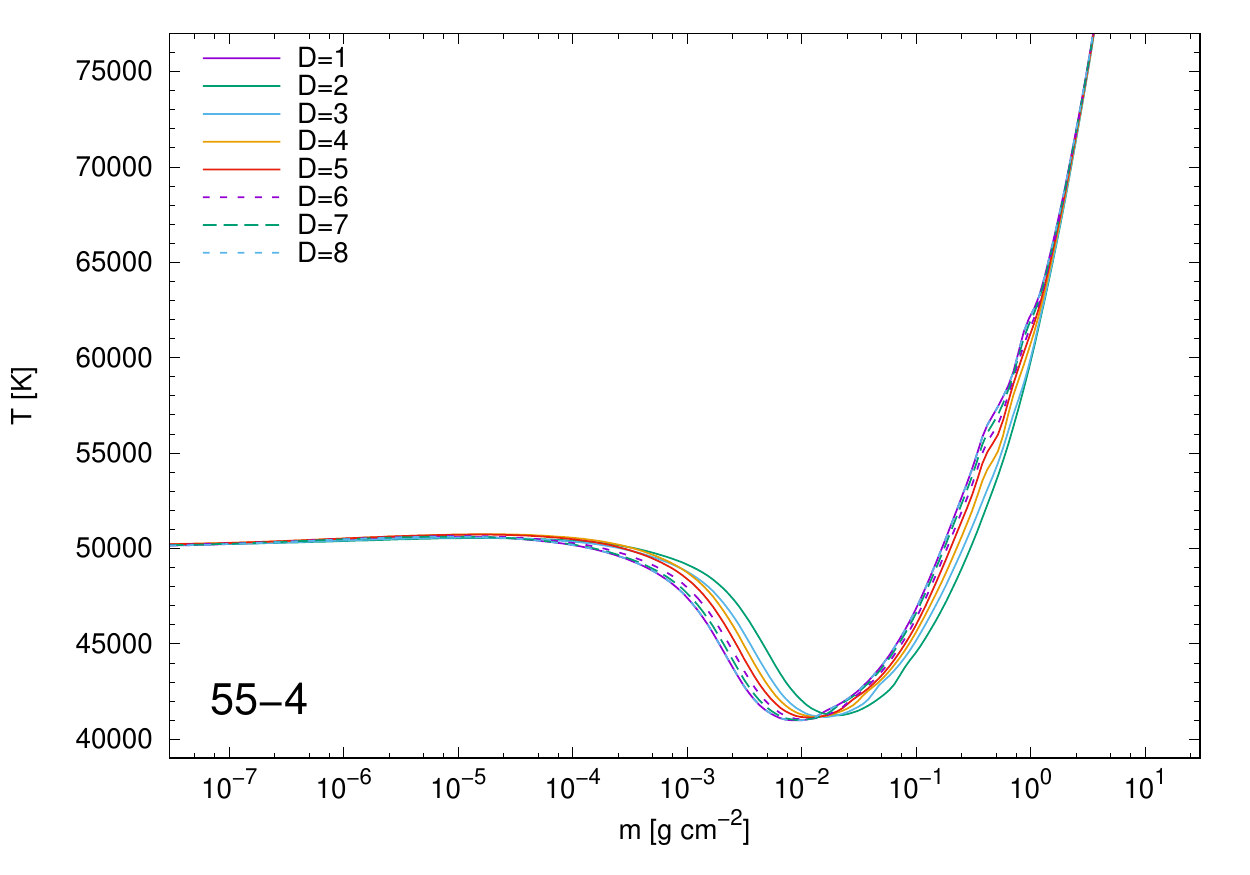}}
\caption{
Same as Fig.~\ref{O6O8B05B2B4B6teplota}, but for models {\sdpetpet}.}
\label{sd55teplota}
\end{figure}

The temperature structure shows a temperature rise at lower optical depths, which is typical for
H-He NLTE model atmospheres \citep[see][Figs.1 -- 4]{Mihalas_Auer_1970}.
A general trend of `shifting the temperature structure outwards' and lowering the temperature
minimum with an increasing clumping factor can be seen in all our model sets
(see Figs. \ref{O6O8B05B2B4B6teplota} and \ref{sd55teplota}).
This is caused by increasing free-free opacity with an increasing clumping factor, which shifts the
continuum formation region upwards to lower $m$.
The temperature in the outermost atmospheric layers does not change significantly with an increasing
clumping factor, especially for hotter atmospheres from our sample.
All changes are within $\sim 1000\,\Kelvin$.
The temperature increase found by our preliminary calculations \citep[Fig.\,1]{statcl} is caused by
the fact that we used mean number densities instead of local ones in thermal balance equations.
The rise is removed by using local (clump) values of number densities.

We also note that for each set of models there is a quite significant difference between the model
without clumping ($D=1$) and that with the clumping factor $D=2$.
This difference from $D=1$ increases with increasing clumping factor; however, differences between
subsequent values of clumping factors from our set ($D=\Dlist$) decrease.

As a consequence of clumping at great depths, there is a significant temperature increase there,
which is caused by increased local Rosseland mean opacity.
As a result of no clumping at great depths, there is no difference between temperature structures at
depths $\tauross\gtrsim1$, as can be seen from Figs. \ref{O6O8B05B2B4B6teplota} and
\ref{sd55teplota}.
The clumping factor was set to~1 for $\tauross\gtrsim 2/3$.

\subsection{Spectral energy distribution}
\label{section_sed}

\begin{figure*}
\resizebox{\hsize}{!}{\includegraphics{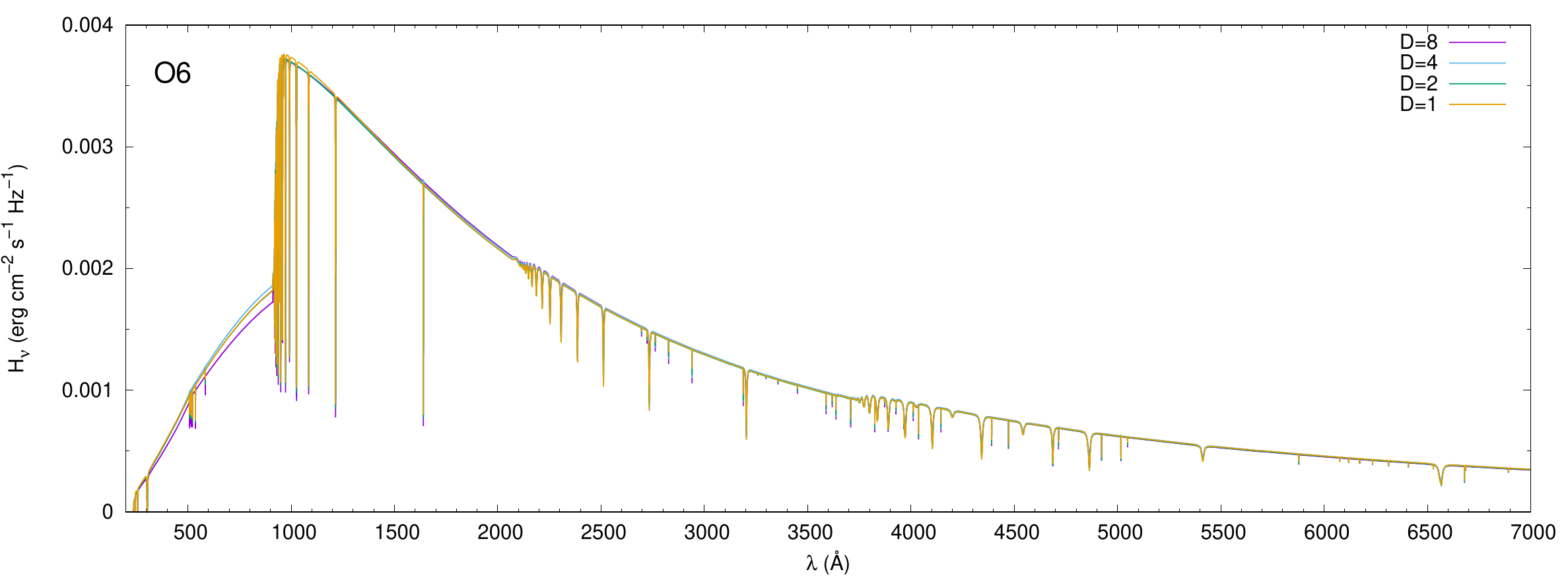}}
\resizebox{\hsize}{!}{\includegraphics{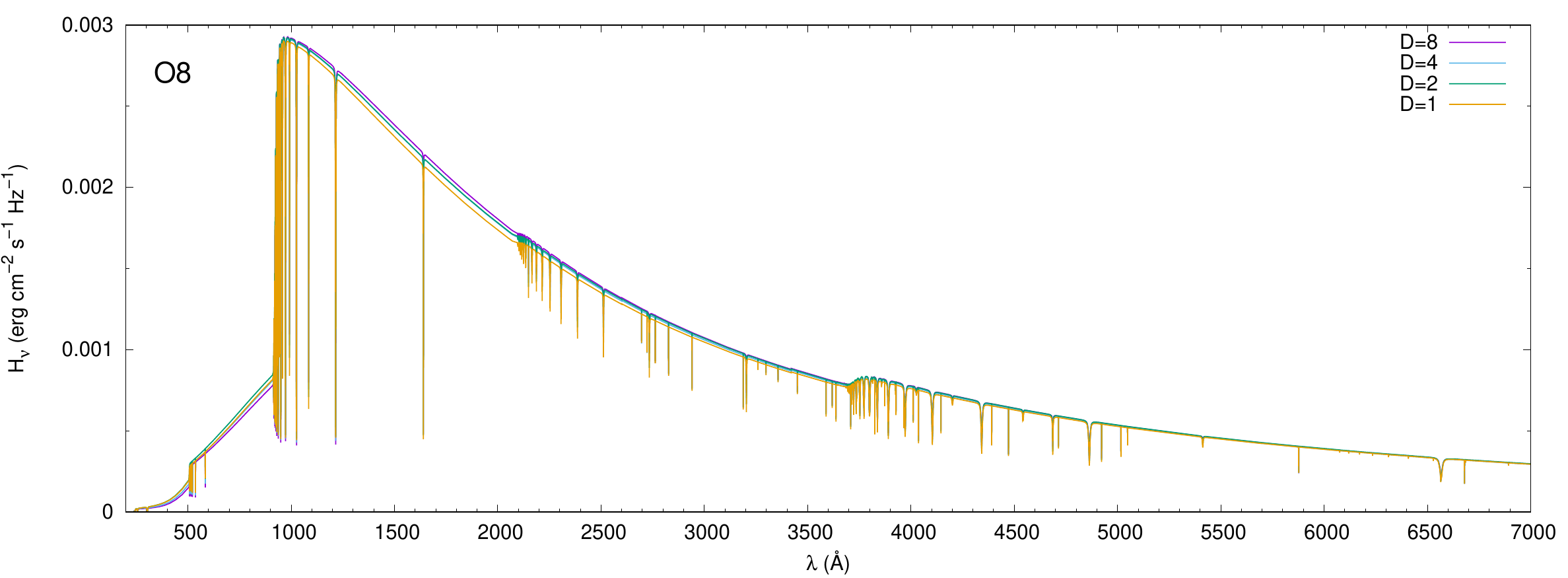}}
\resizebox{\hsize}{!}{\includegraphics{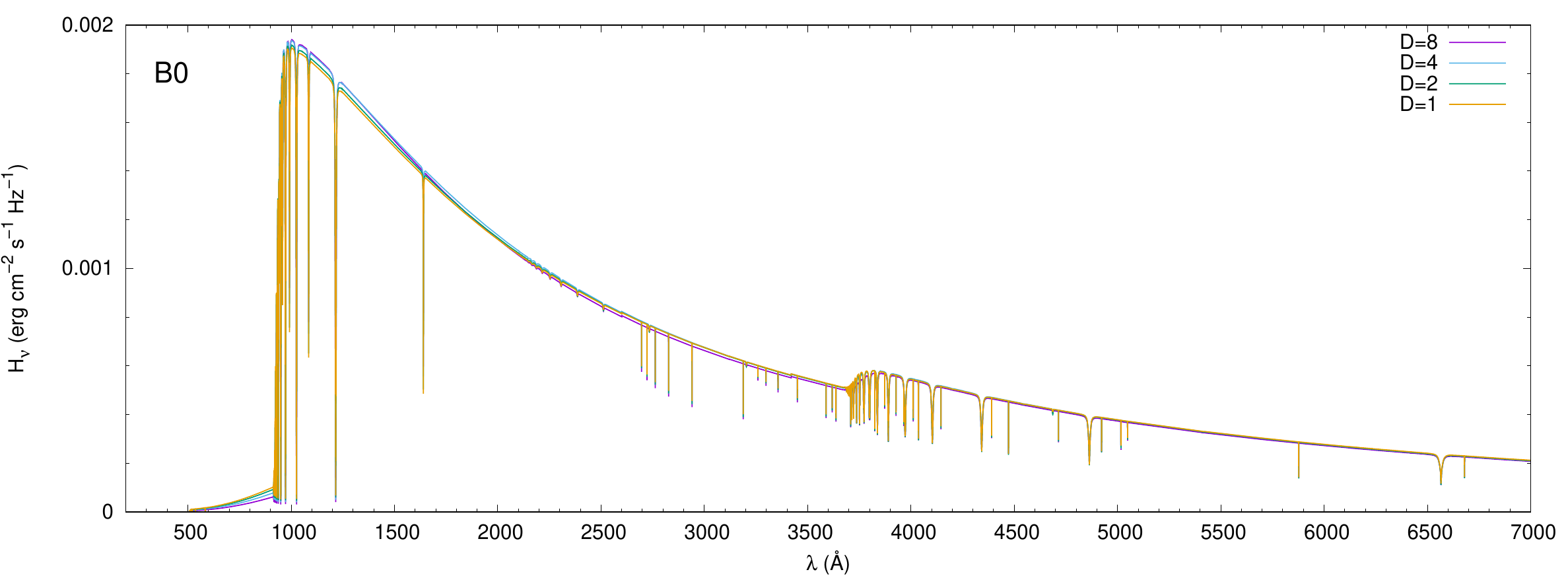}}
\caption{UV and visual flux for main-sequence models {\Osest} (upper panel), {\Oosm} (middle panel),
and {\Bnula} (lower panel).
For global model parameters, see Table~\ref{parametry_modelu}.}
\label{O6O8B05flux}
\end{figure*}
\begin{figure*}
\resizebox{\hsize}{!}{\includegraphics{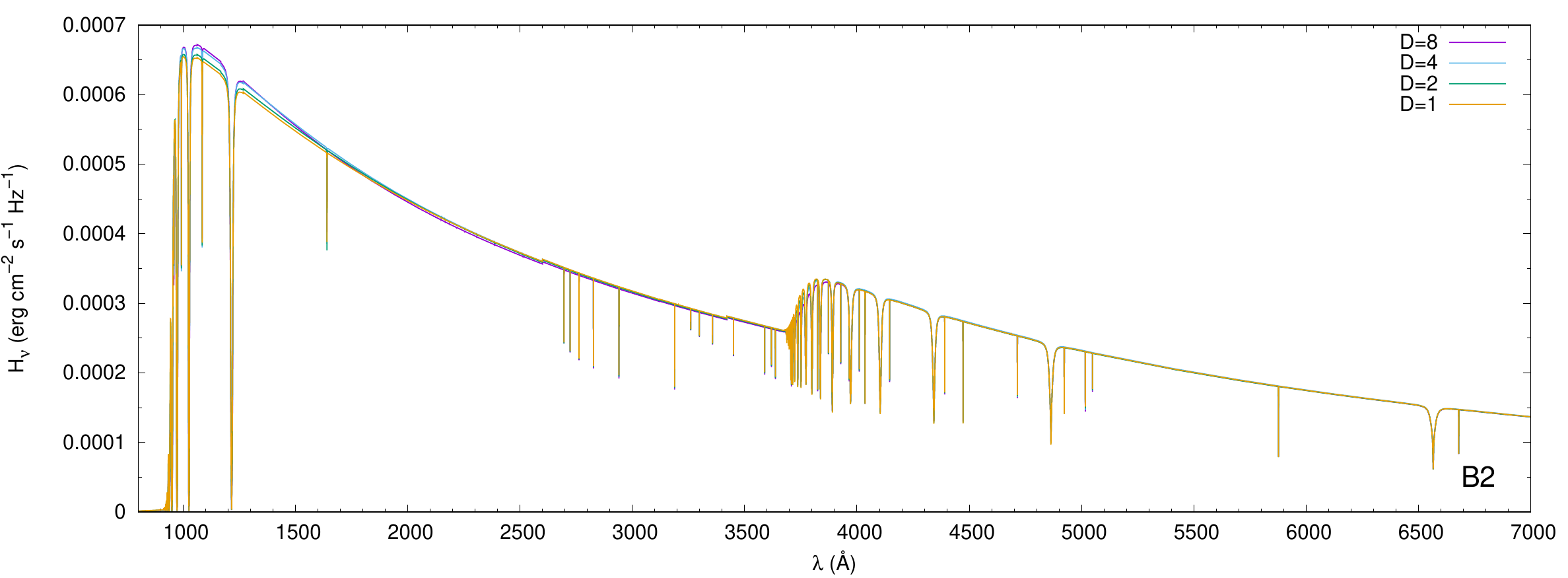}}
\resizebox{\hsize}{!}{\includegraphics{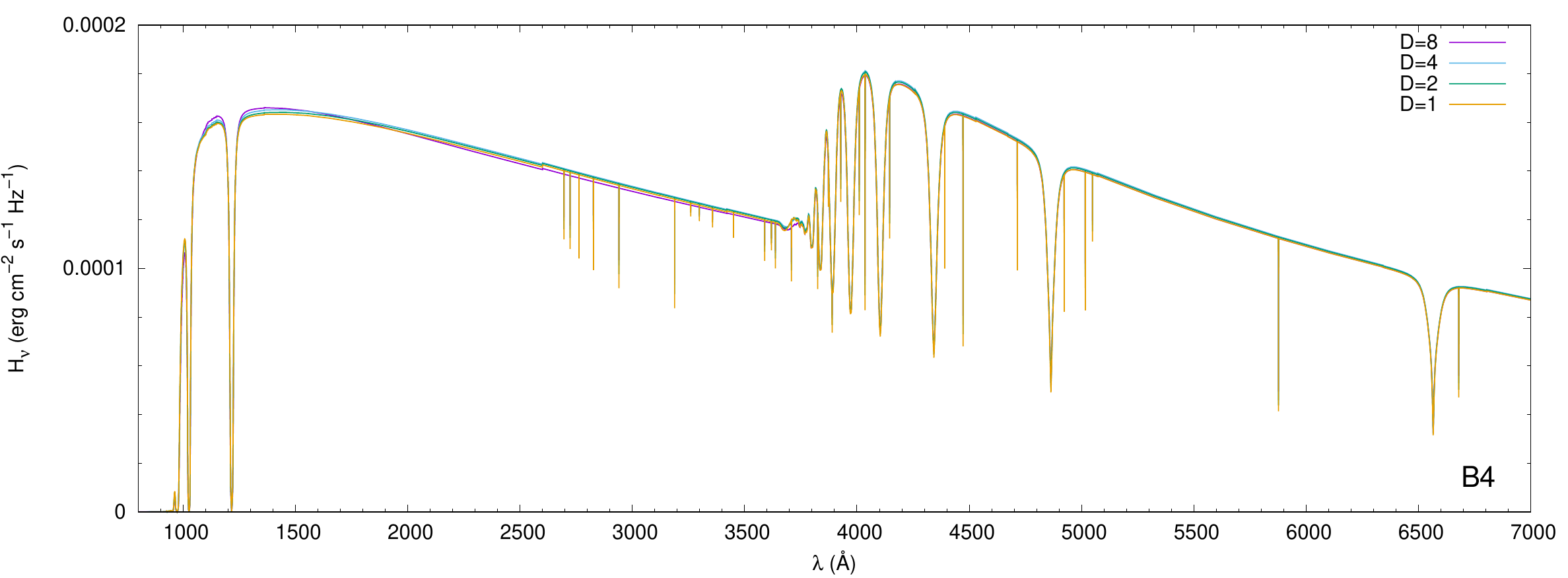}}
\resizebox{\hsize}{!}{\includegraphics{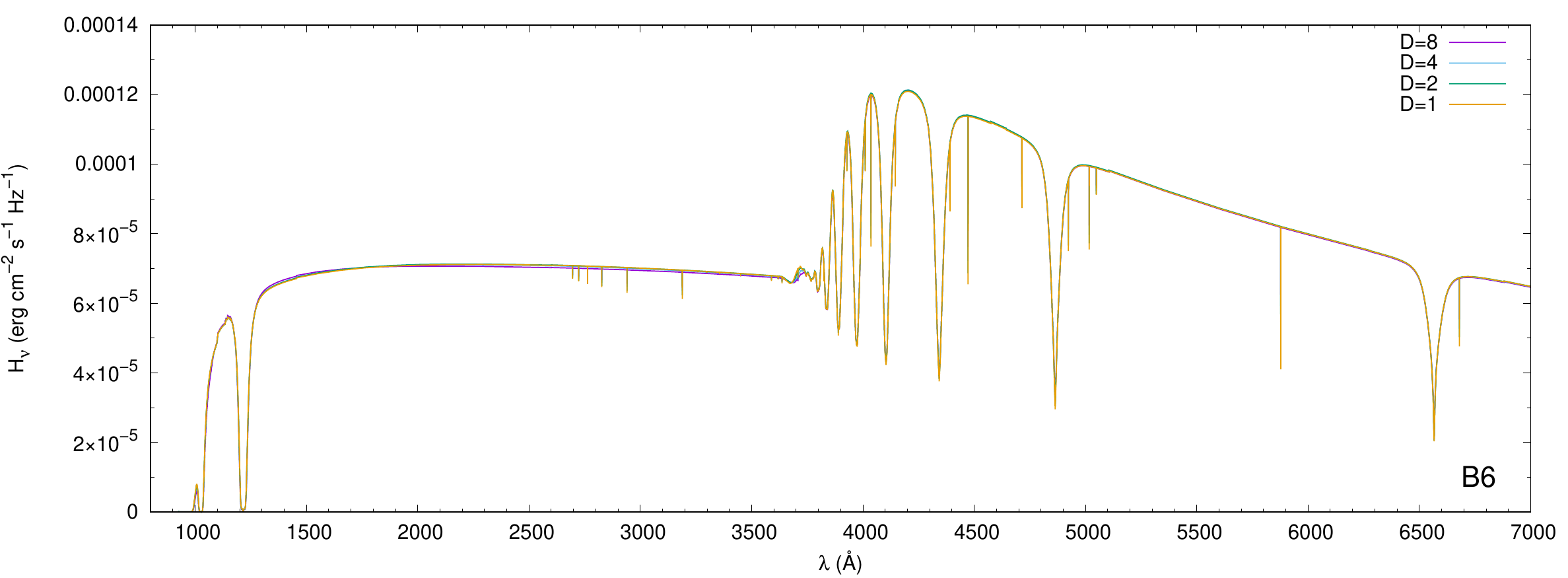}}
\caption{Same as Fig.~\ref{O6O8B05flux}, but for main-sequence models {\Bdva} (upper panel),
{\Bctyri} (middle panel), and {\Bsest} (lower panel).}
\label{B2B4B6flux}
\end{figure*}

\begin{figure*}
\resizebox{\hsize}{!}{\includegraphics{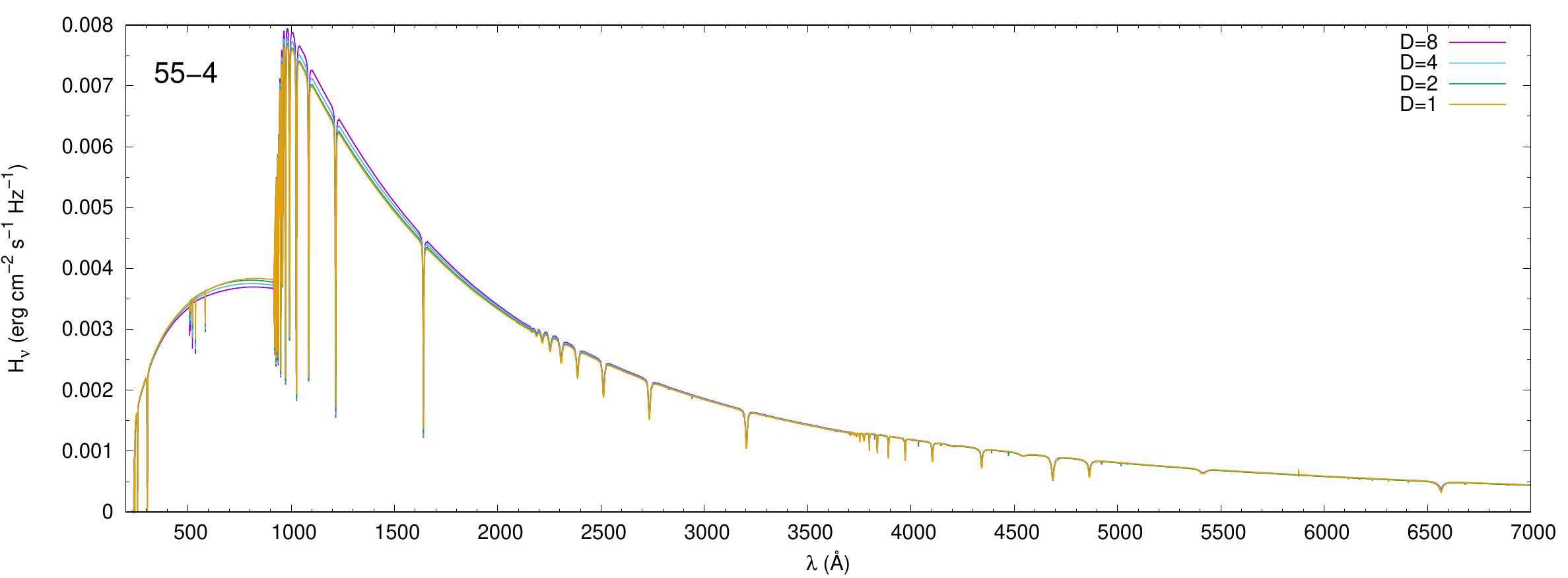}}
\caption{Same as Fig.~\ref{O6O8B05flux}, but for subdwarf models {\sdpetpet}.}
\label{sd55flux}
\end{figure*}

The most interesting observable effect of including a clumping factor in model atmosphere
calculation is the change of the emergent radiation flux (spectral energy distribution).
The effect is illustrated in Figures \ref{O6O8B05flux} and \ref{B2B4B6flux} for clumping factors
$\clfact=1,2,4$, and $8$.
Clumping in the model atmosphere causes an enhancement of the emergent flux in the part of the UV
spectral region where the hydrogen Lyman series lines form.
This is accompanied by a lowering of the flux in other spectral regions, depending on the specific
model.
Maximum changes are for the hottest star model from our sample ({\sdpetpet, Fig.\,\ref{sd55flux}),
the flux near the hydrogen Lyman ionisation edge for the model with $\clfact=8$ is about $5\%$
larger for frequencies below the ionisation edge than the model without clumping and about $5\%$
smaller for frequencies above the ionisation edge.
The Lyman series flux increases with an increasing clumping factor.
In the visual region, fluxes for clumped and unclumped models are roughly the same, and in the
infrared region clumped models have lower radiation fluxes.
This behaviour is opposite to the flattening of the continuum that is typical of a transfer
from plane-parallel to spherically symmetric model atmospheres \citep[see][]{Kunasz_etal_1975,
Gruschinske_Kudritzki_1979, spcspn}.
Direct enhancement of the opacity by clumping is just one of the reasons for this behaviour,
since only the free-free opacity is affected (enhanced) by clumping, all ionisation and line
opacities are the same both for clumped and unclumped models.
The emissivity is affected similarly, free-bound and free-free emissivities are $D$-times enhanced
(for all wavelengths), and line emissivity is not directly affected.
However, the opacities and emissivities are indirectly affected by changes in temperature and
density structure as well as by changes in the atomic level number densities.
As the corresponding equations describing the physical processes included (radiative transfer
equation, hydrostatic equilibrium equation, radiative equilibrium or thermal balance equation, and
the kinetic equilibrium equations) couple all substantial quantities, all coupling among physical
processes is taken into account and the equations are solved simultaneously.
Consequently, it is impossible to pick one single process responsible for the changes in emergent
radiation caused by clumping.

Similar effects on the spectral energy distributions can also be found for other sets of models from
our sample (see Figs. \ref{O6O8B05flux} and \ref{B2B4B6flux}).
All models display flux enhancement in some part of the UV region and a decrease in other parts
of the spectrum with an increasing clumping factor; this effect is weaker with a lower stellar
effective temperature.

\begin{figure}
\centering
\resizebox{\hsize}{!}{\includegraphics{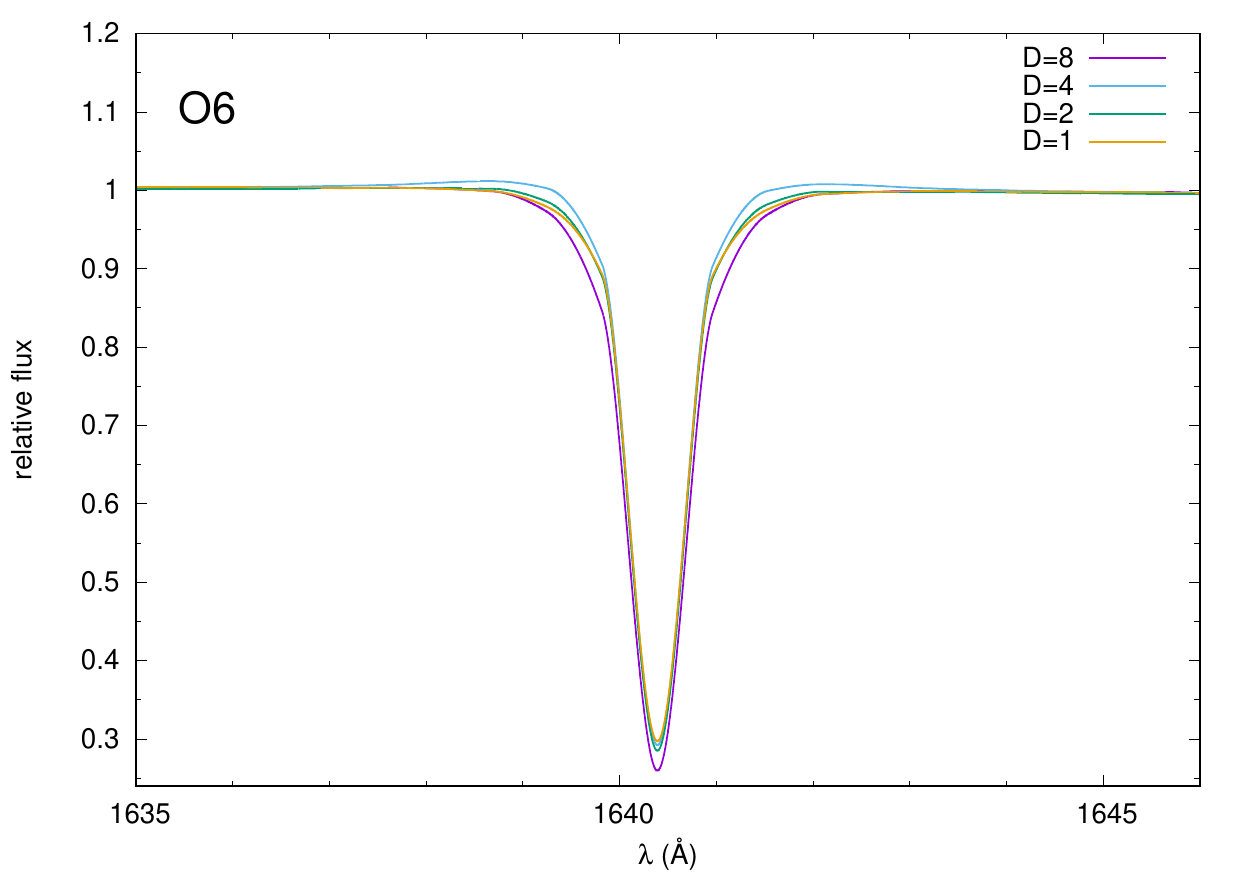}}
\resizebox{\hsize}{!}{\includegraphics{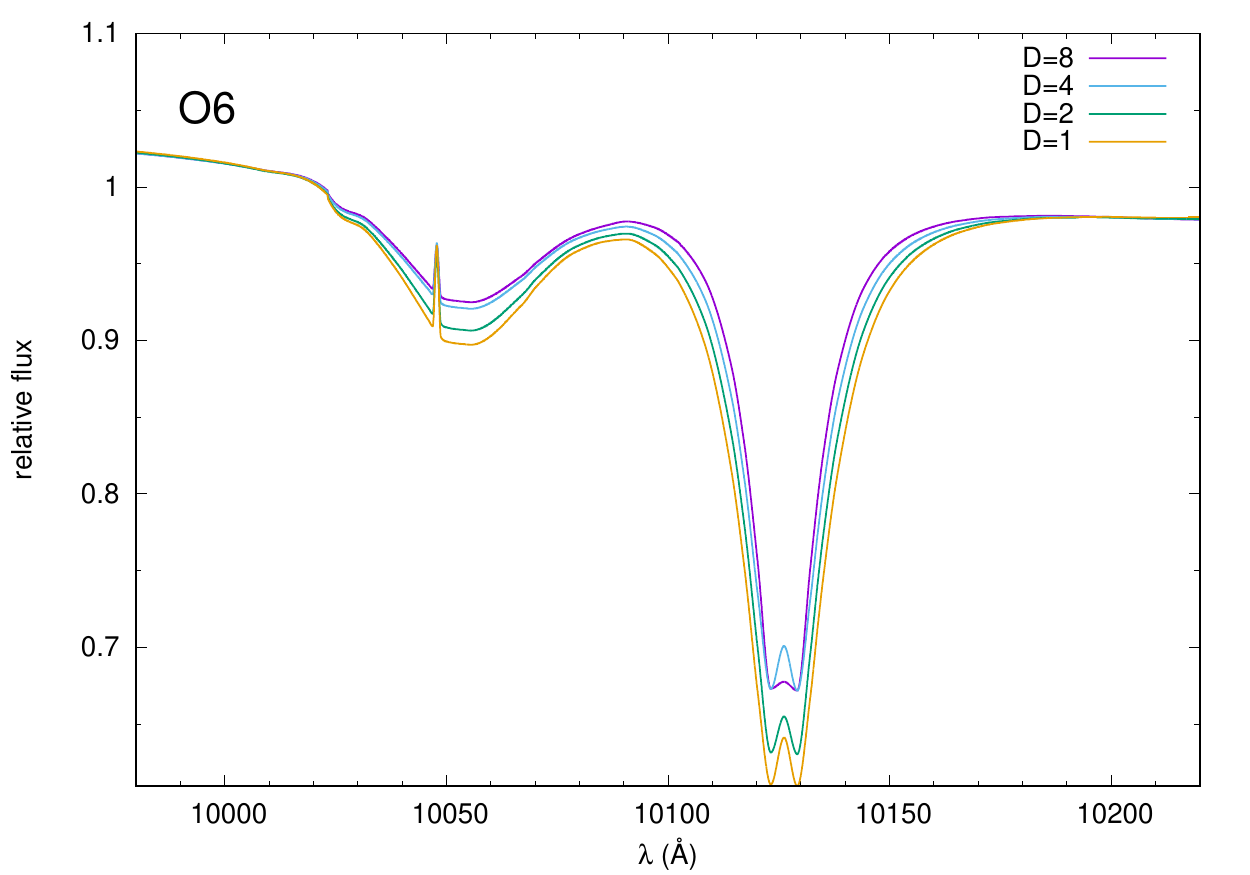}}
\caption{Selected normalised line profiles for the {\Osest} model for clumping factors $\clfact=1,
2, 4, 8$.
\emph{Upper panel:} \ion{He}{ii} {\Halpha} line profile.
\emph{Lower panel:} \ion{He}{ii} {\Pialpha} (4-6, 10123\AA) and \ion{H}{i} {\Pdelta} line  with
an emission peak of the \ion{He}{ii} 6-14 line (10048{\AA}) line profiles.}
\label{O6-lines}
\end{figure}

Selected spectral lines also show differences between clumped and unclumped models.
The sensitivity of lines with respect to clumping depends on the corresponding line formation depth
and its changes with clumping.
Since the line formation temperatures and densities may change in both directions, some lines are
fainter with increasing clumping factor, and some other lines become stronger.
This can be illustrated for the {\Osest} model.
When expressed in intensities related to continuum, the absorption \ion{He}{ii} {\Halpha} line at
$1640.5\,\AA$ is stronger for a larger clumping factor, while the \ion{He}{ii} {\Pialpha} line is
weaker for a larger clumping factor (see Fig.\,\ref{O6-lines}).
The latter figure shows a similar effect for the {\Pdelta} line.
All changes are of the order of \% in line centres.
Due to a simplified treatment of the chemical composition in our models, our plots show only
hydrogen and helium lines.

\subsection{Non-equilibrium atomic level populations}

\begin{figure}
\centering
\resizebox{\hsize}{!}{\includegraphics{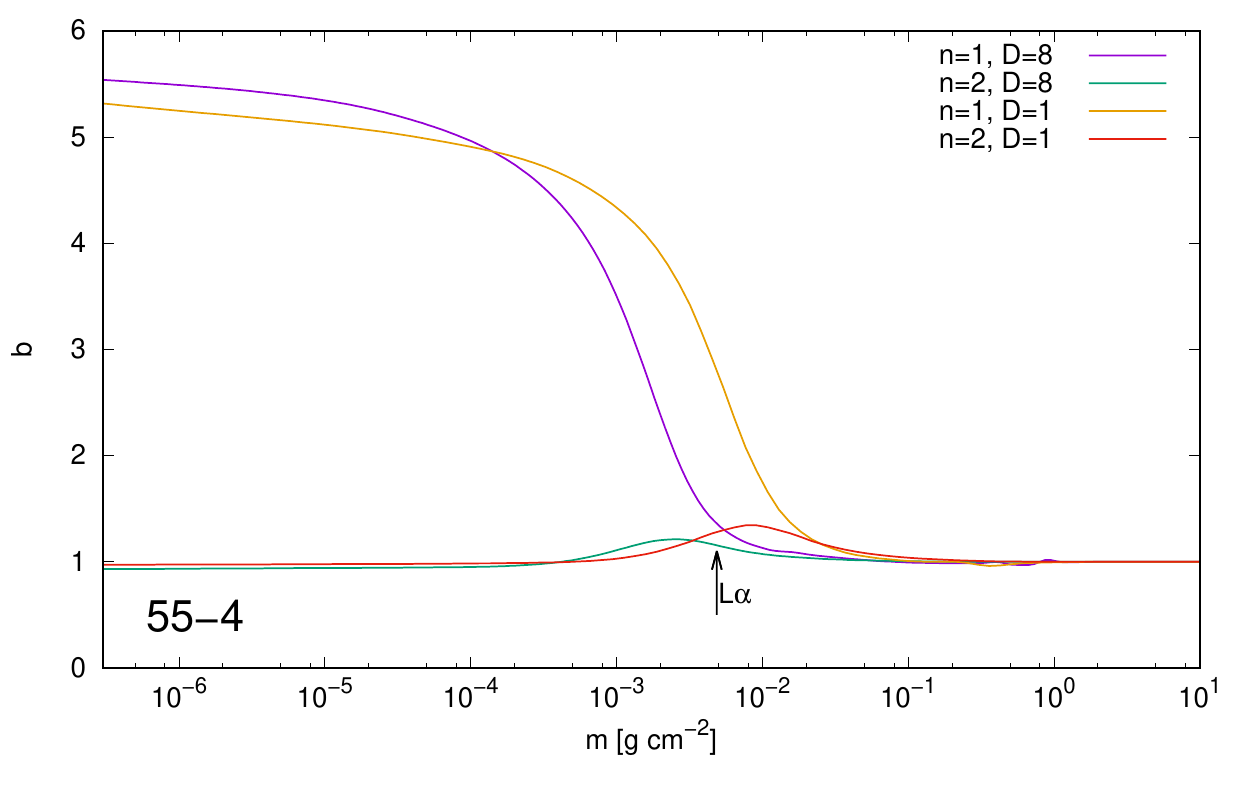}}
\resizebox{\hsize}{!}{\includegraphics{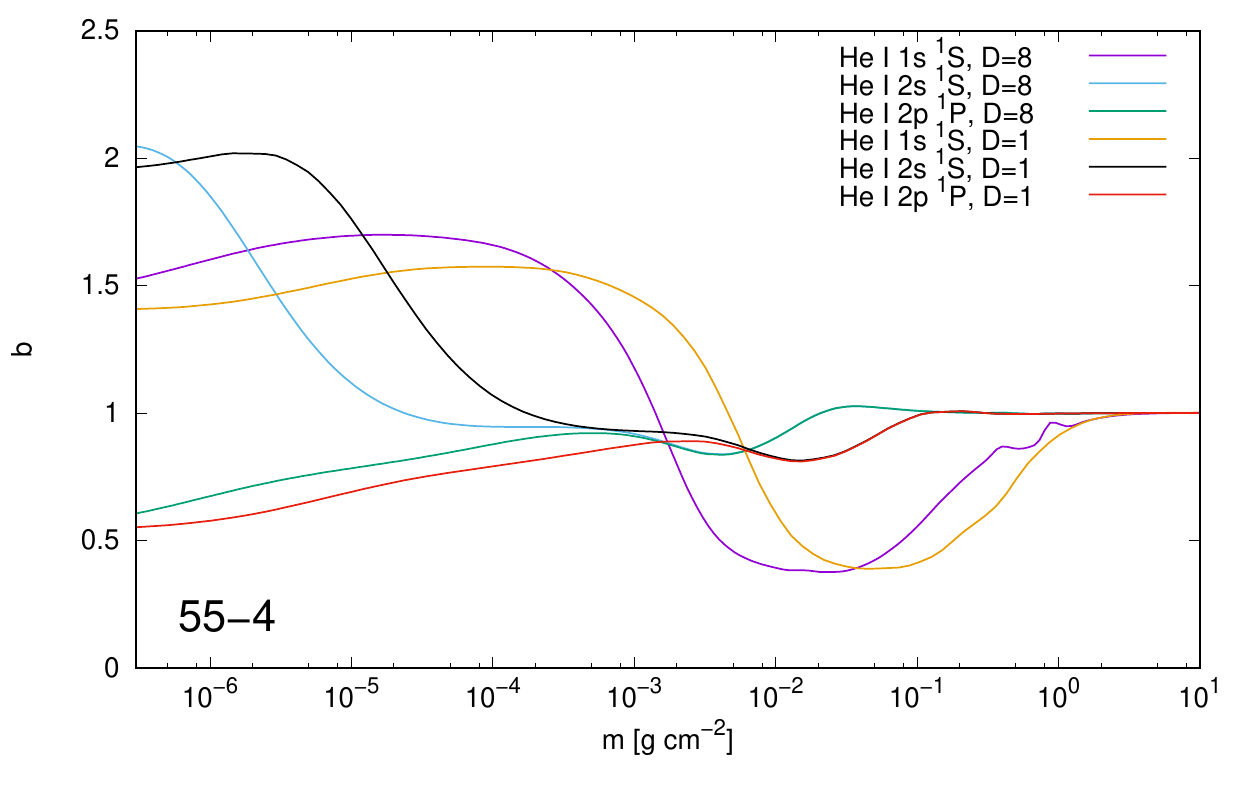}}
\resizebox{\hsize}{!}{\includegraphics{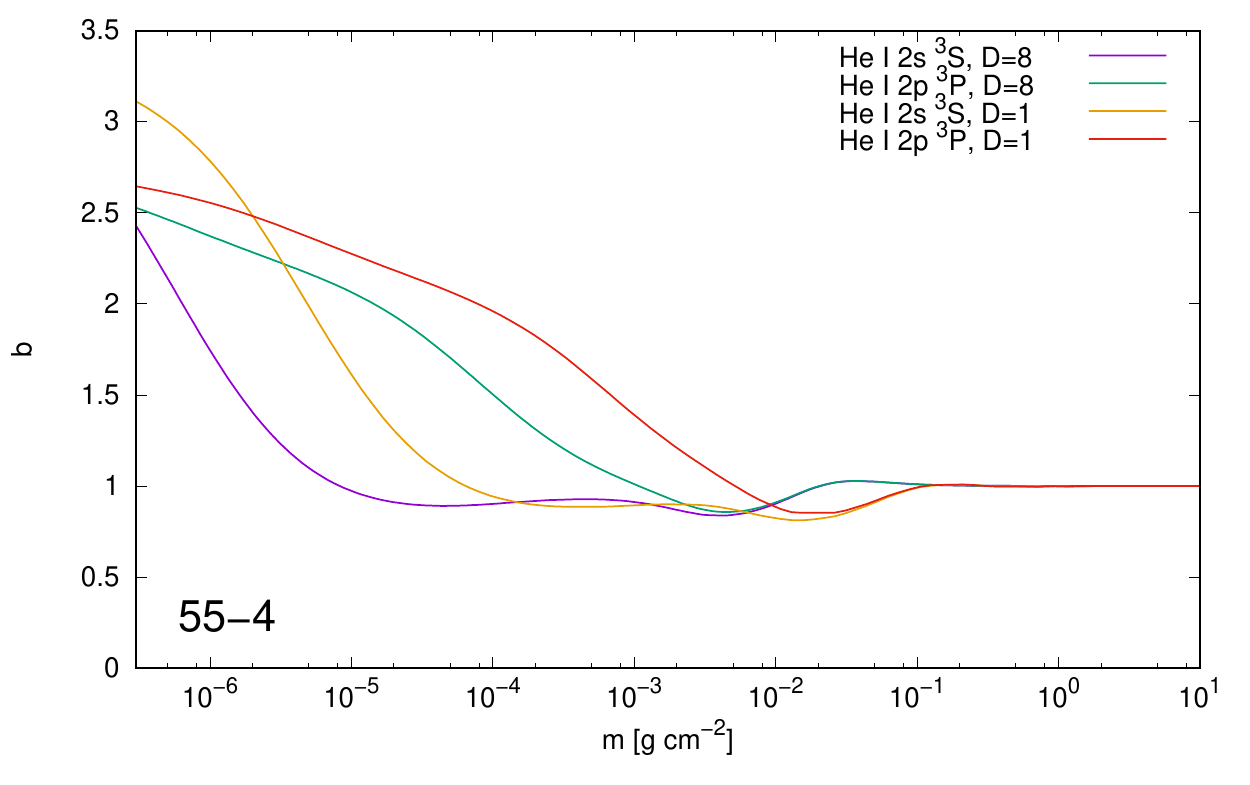}}
\resizebox{\hsize}{!}{\includegraphics{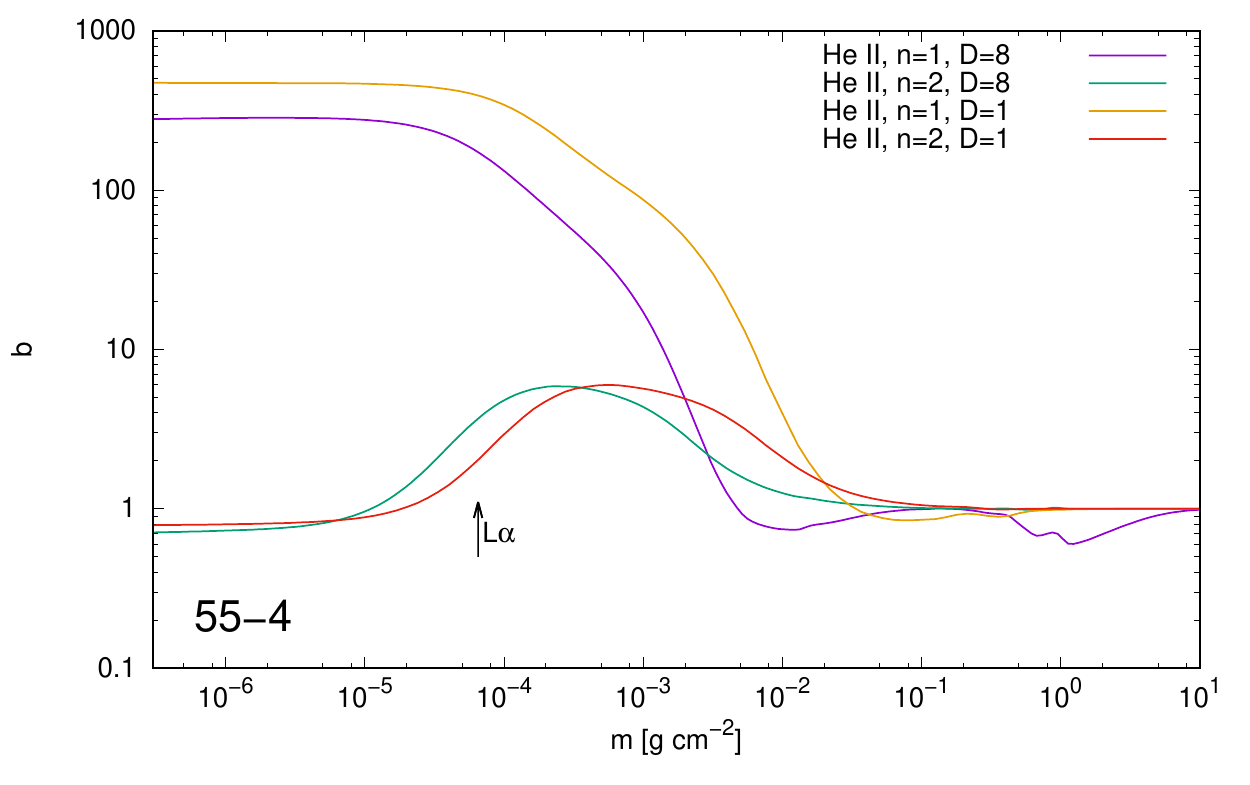}}
\caption{
Departure coefficients ($b$-factors) for clumping factors $\clfact=1$ and $\clfact=8$ and for
selected atomic levels for a set of model atmospheres {\sdpetpet} (for basic global model
parameters, see Table\,\ref{parametry_modelu}).
From the top panel:
\ion{H}{i} $n=1$ and $n=2$ levels;
\ion{He}{i} $1s\,\termls{1}{S}$, $2s\,\termls{1}{S}$, and $2p\,\termls{1}{P}$ levels;
\ion{He}{i} $2s\,\termls{3}{S}$ and $2p\,\termls{3}{P}$ levels;
\ion{He}{ii} $n=1$ and $n=2$ levels.
}
\label{55bfac}
\end{figure}

\begin{figure}
\centering
\resizebox{\hsize}{!}{\includegraphics{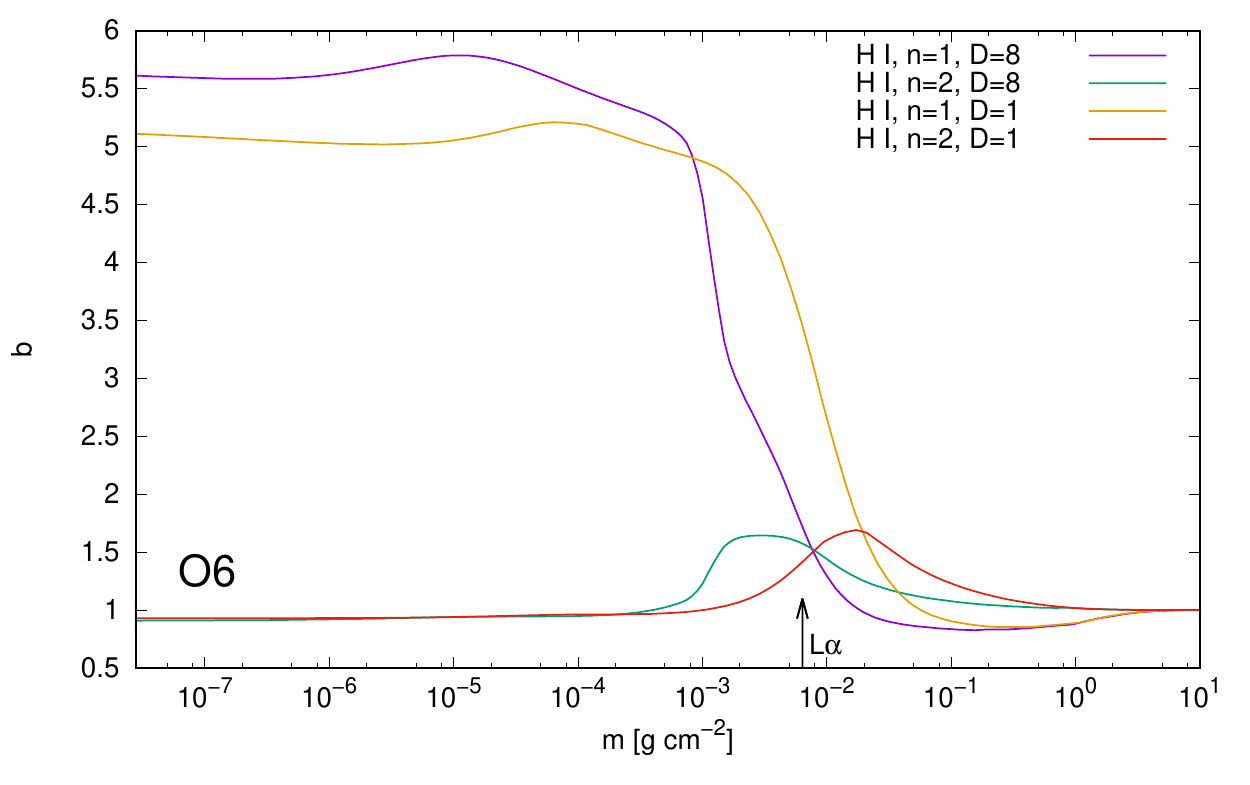}}
\resizebox{\hsize}{!}{\includegraphics{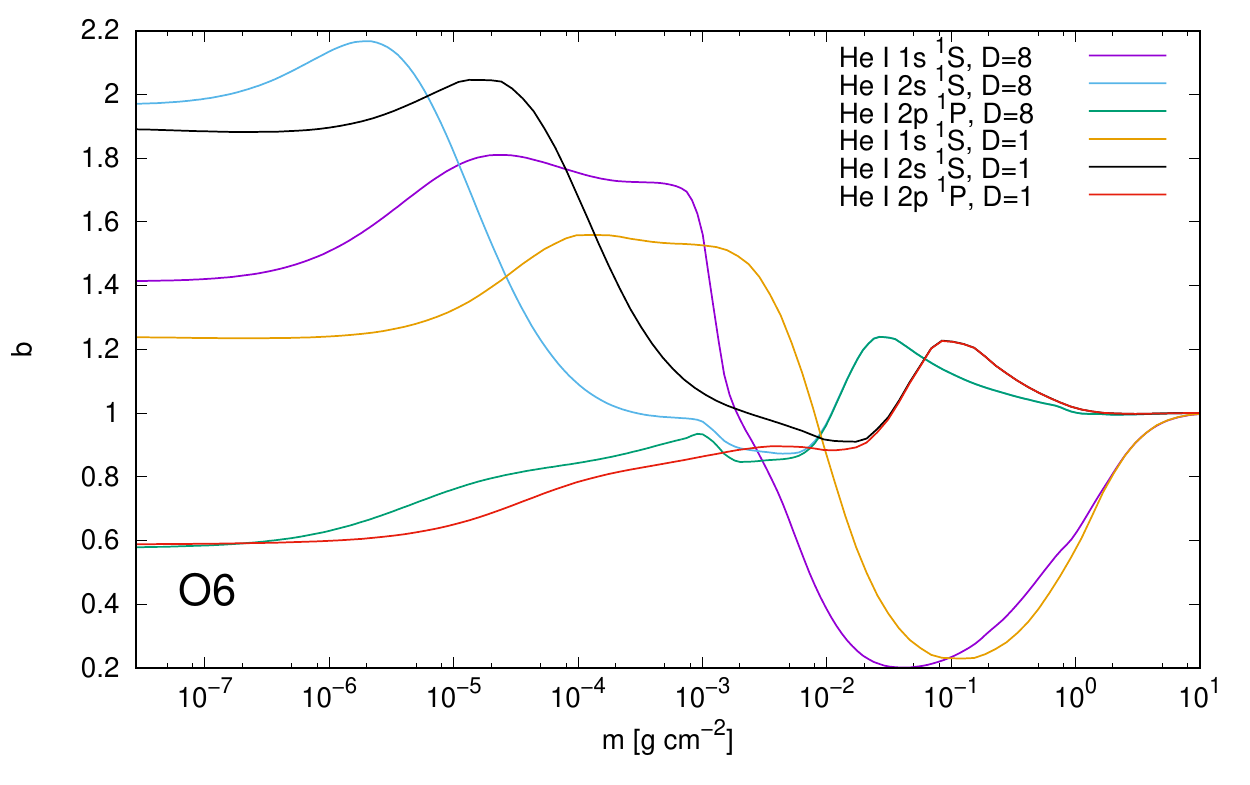}}
\resizebox{\hsize}{!}{\includegraphics{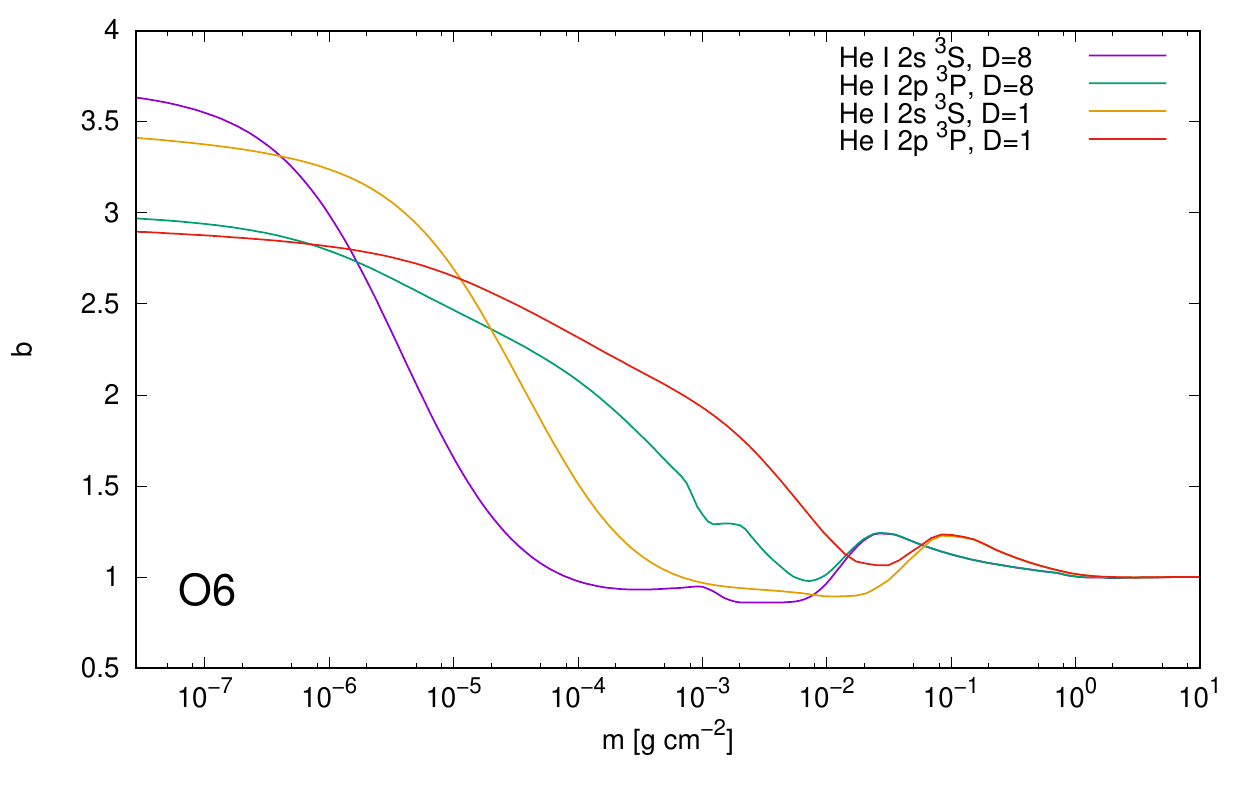}}
\resizebox{\hsize}{!}{\includegraphics{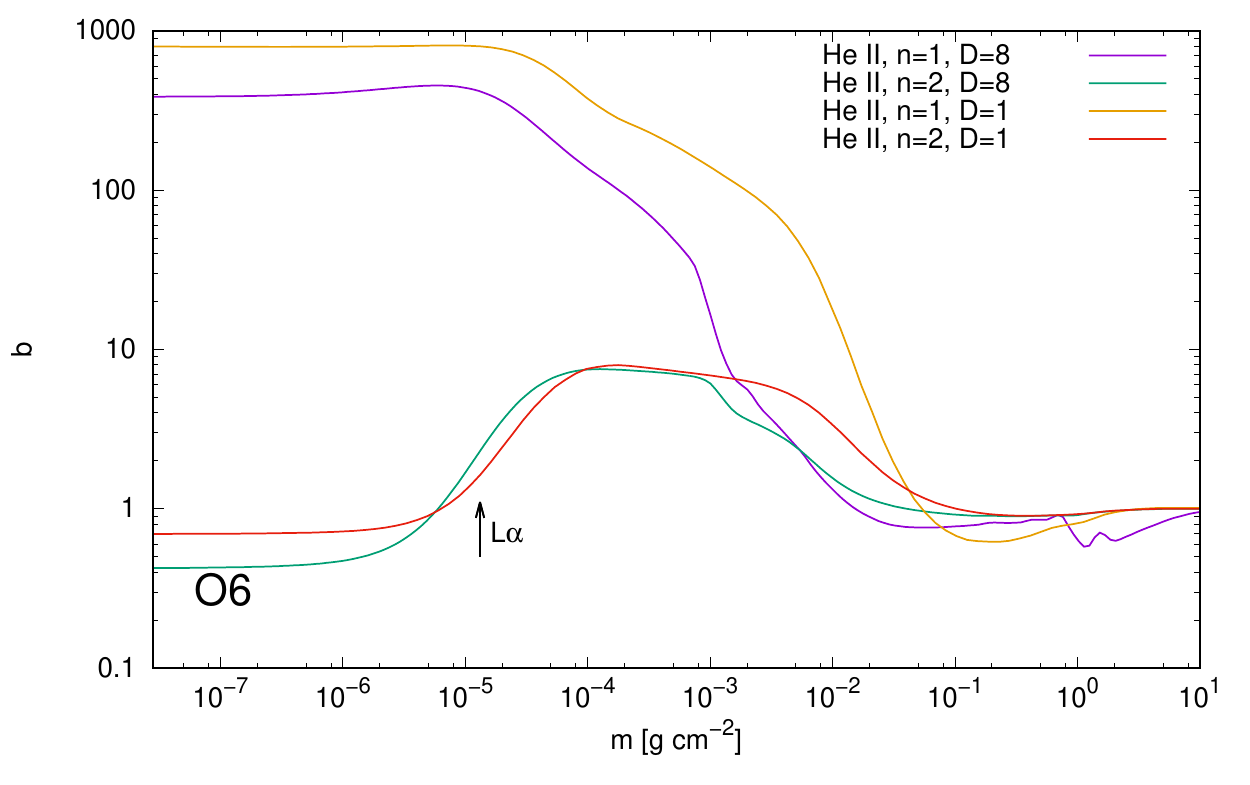}}
\caption{Same as Fig.\,\ref{55bfac}, but for the model set {\Osest}.}
\label{O6bfac}
\end{figure}

\begin{figure}
\centering
\resizebox{\hsize}{!}{\includegraphics{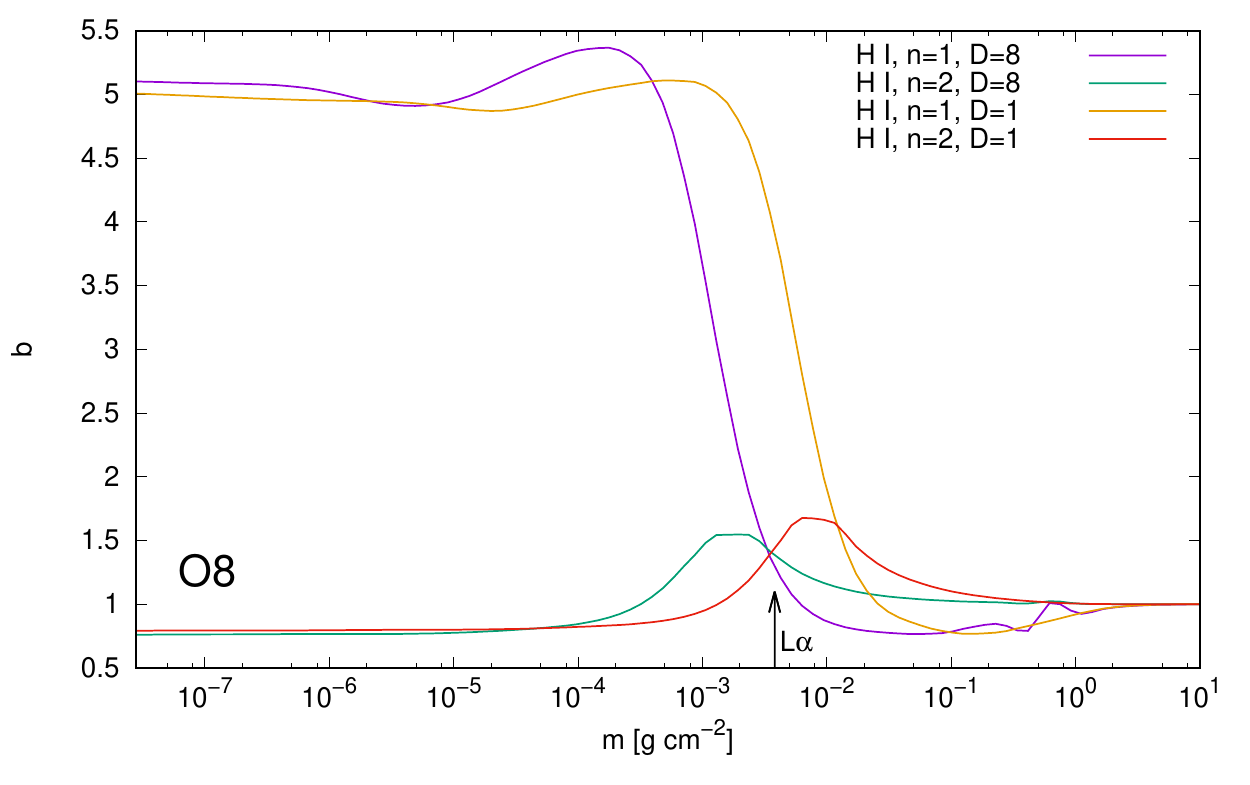}}
\resizebox{\hsize}{!}{\includegraphics{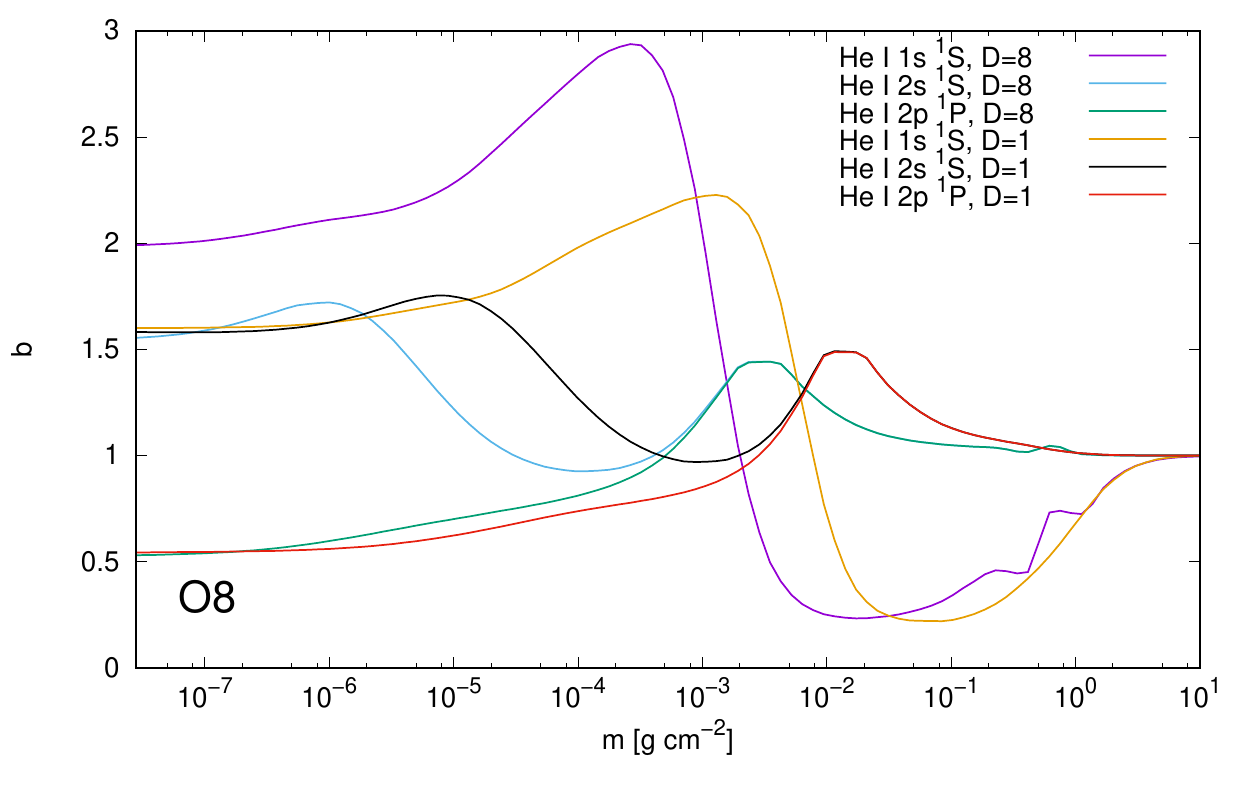}}
\resizebox{\hsize}{!}{\includegraphics{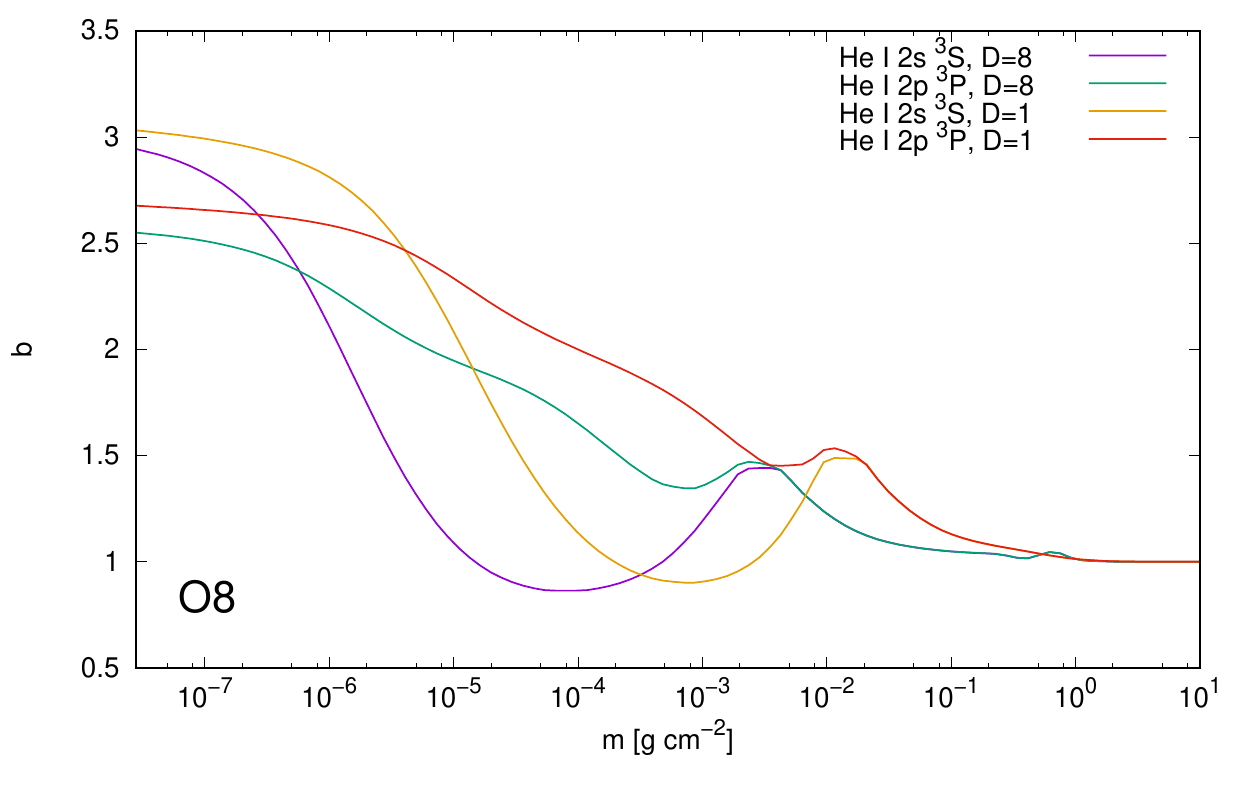}}
\resizebox{\hsize}{!}{\includegraphics{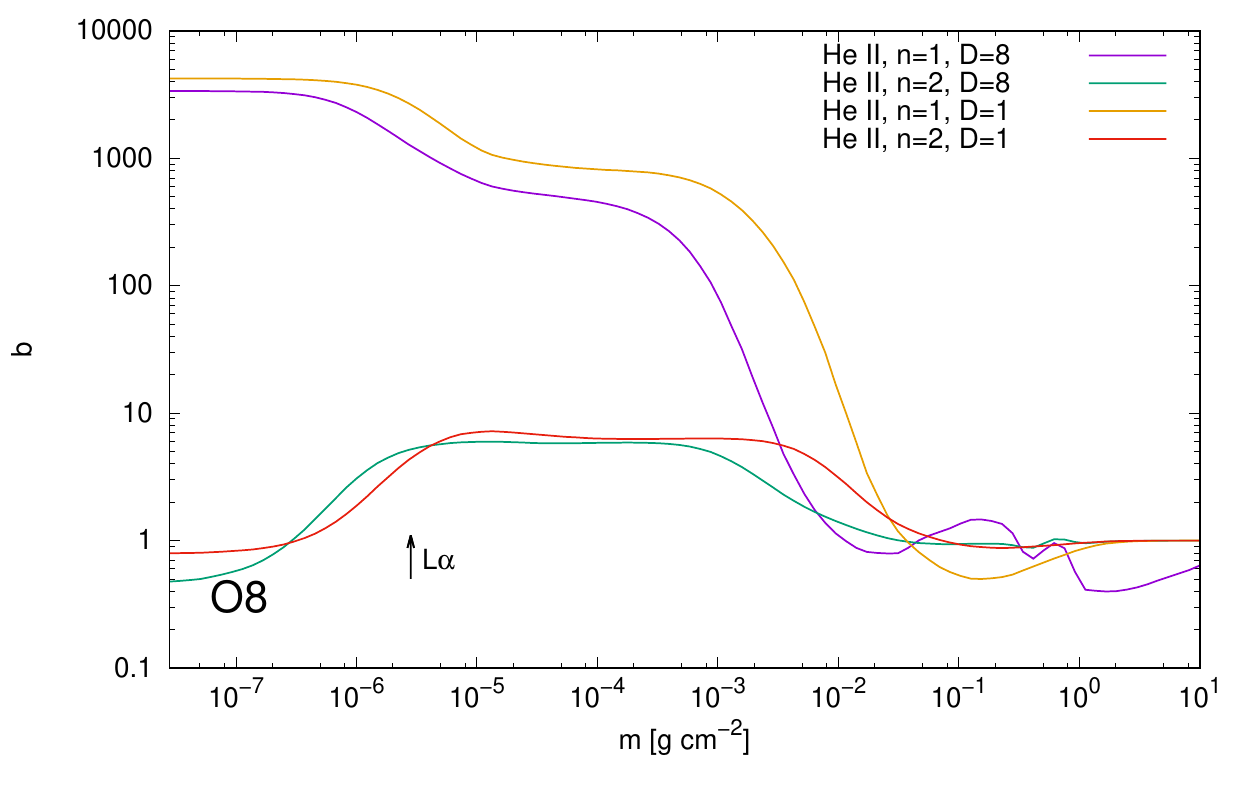}}
\caption{Same as Fig.\,\ref{55bfac}, but for the model set {\Oosm}.}
\label{O8bfac}
\end{figure}

\begin{figure}
\centering
\resizebox{\hsize}{!}{\includegraphics{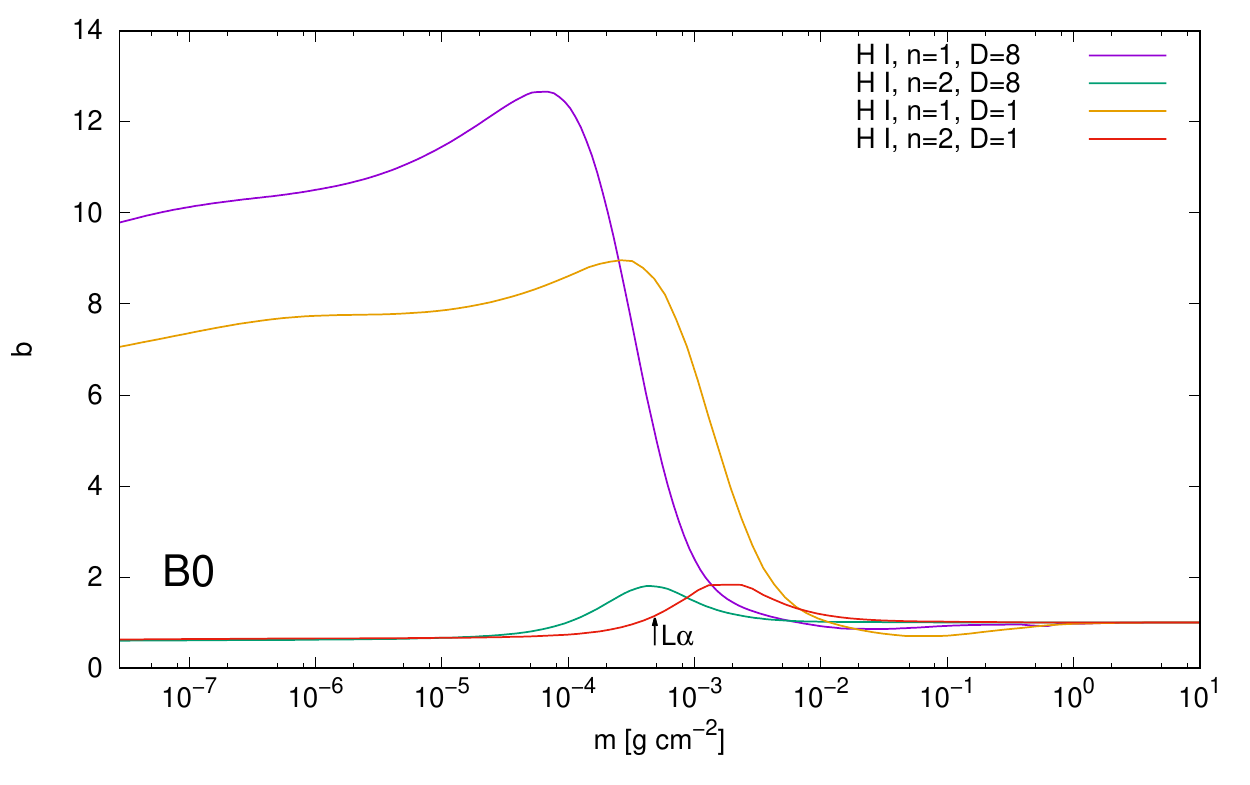}}
\resizebox{\hsize}{!}{\includegraphics{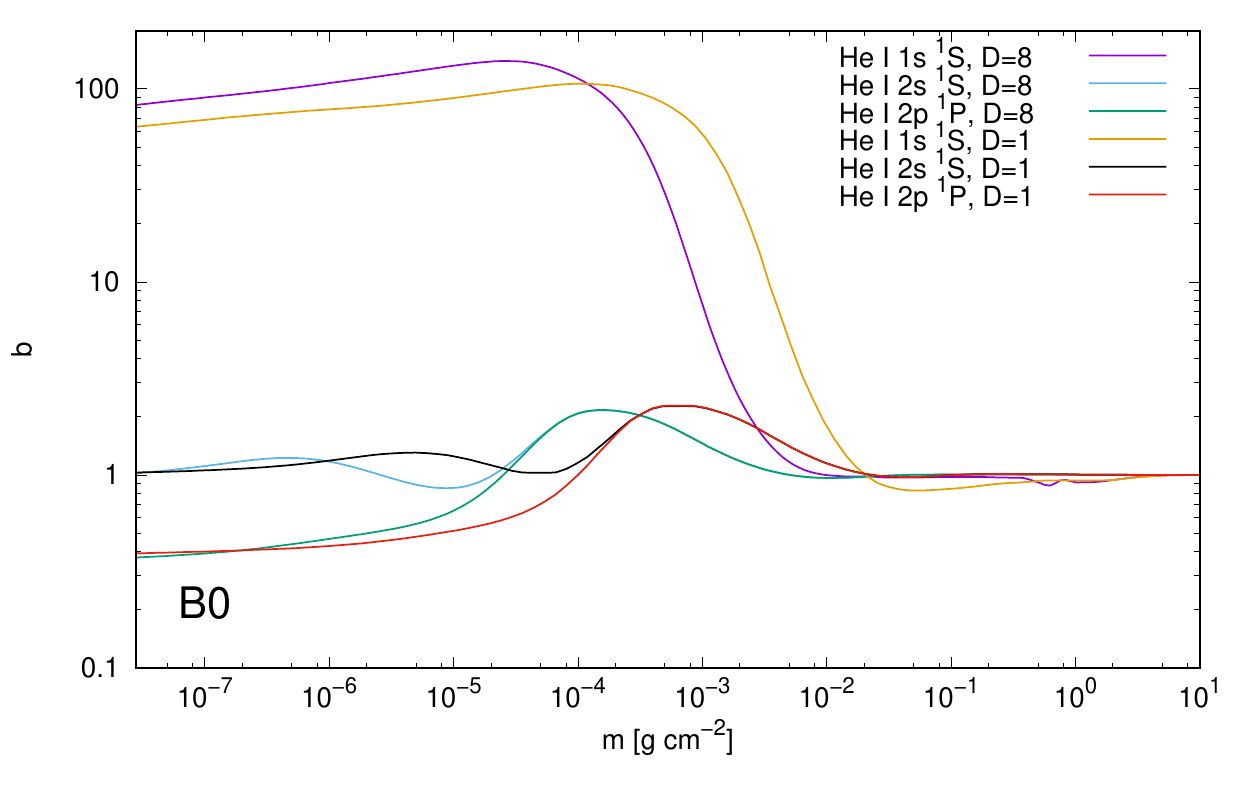}}
\resizebox{\hsize}{!}{\includegraphics{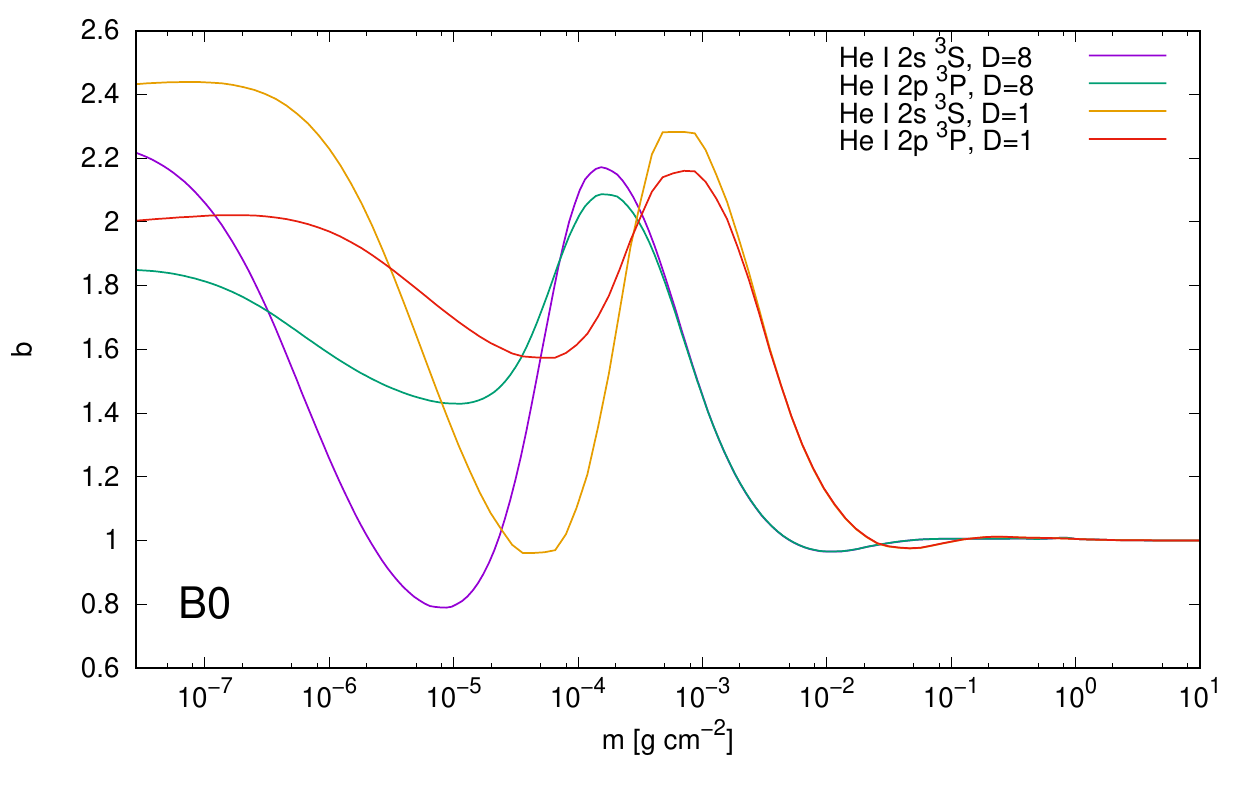}}
\resizebox{\hsize}{!}{\includegraphics{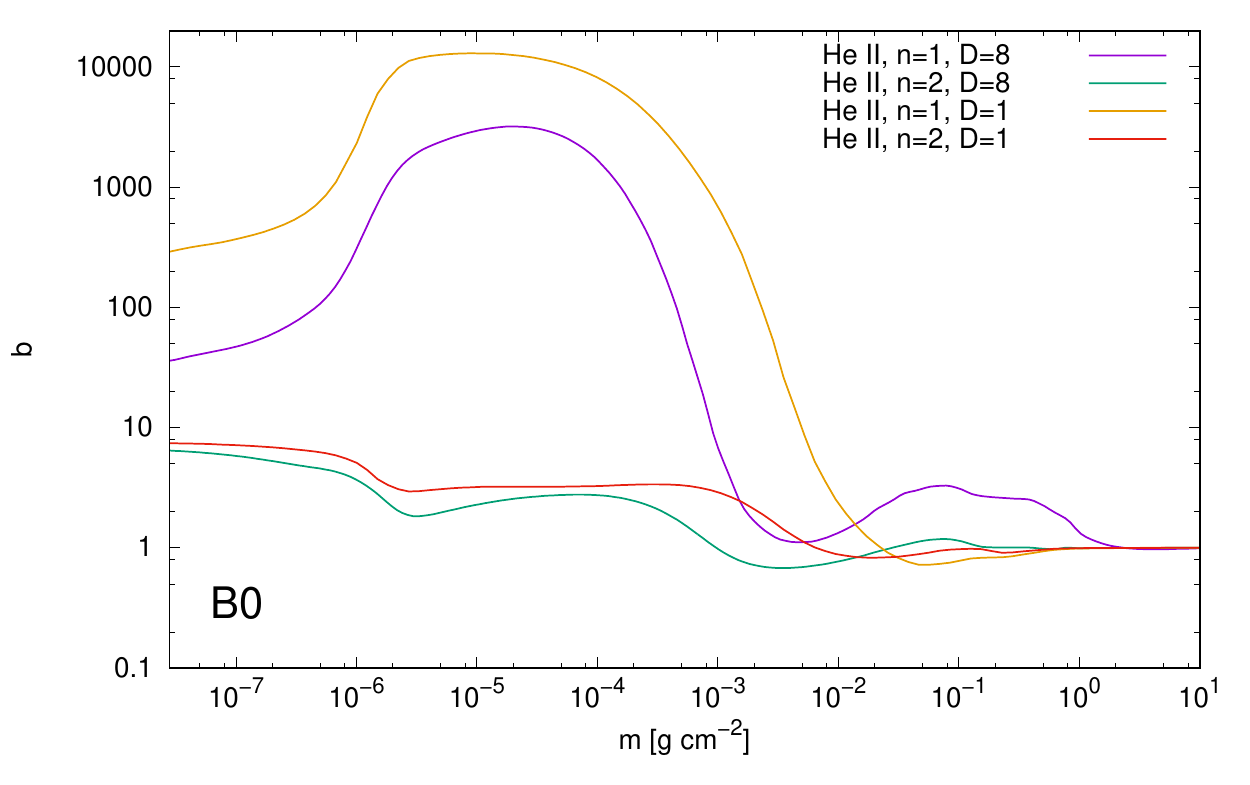}}
\caption{Same as Fig.\,\ref{55bfac}, but for the model set {\Bnula}.}
\label{B0bfac}
\end{figure}

\begin{figure}
\centering
\resizebox{\hsize}{!}{\includegraphics{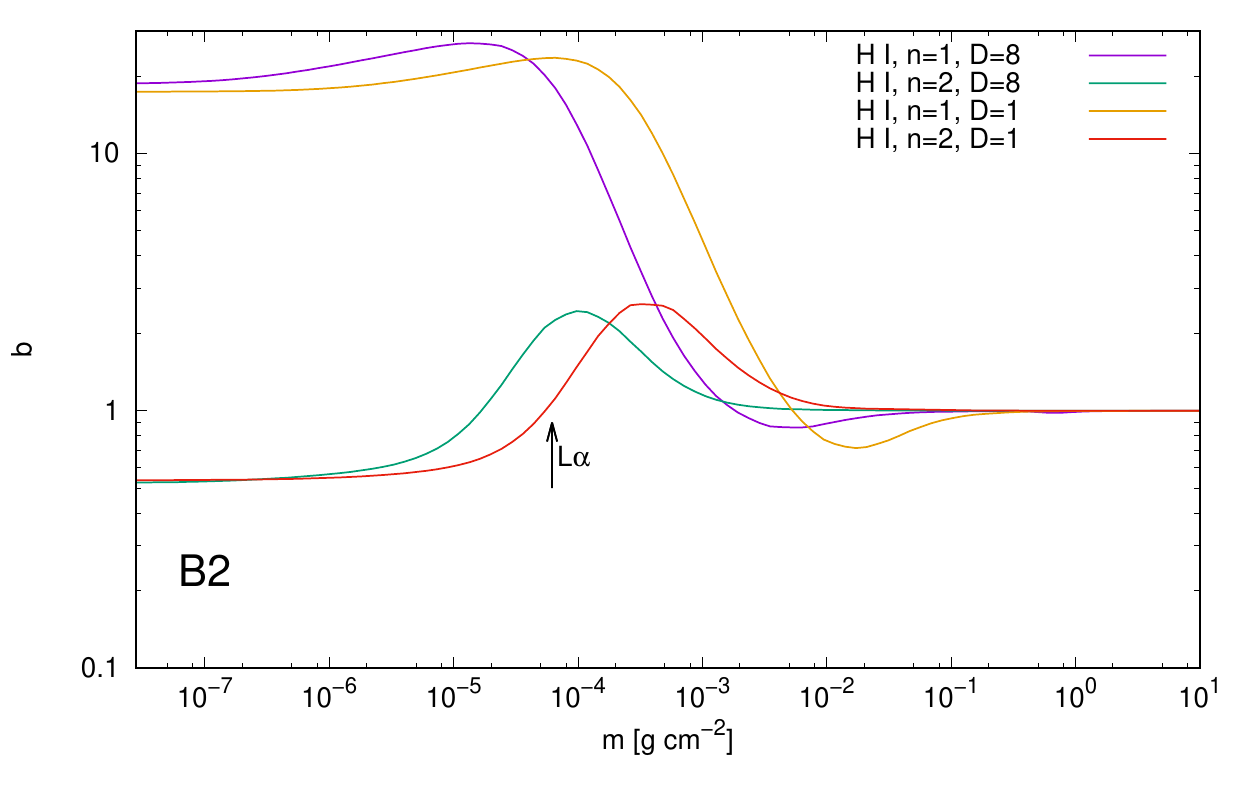}}
\resizebox{\hsize}{!}{\includegraphics{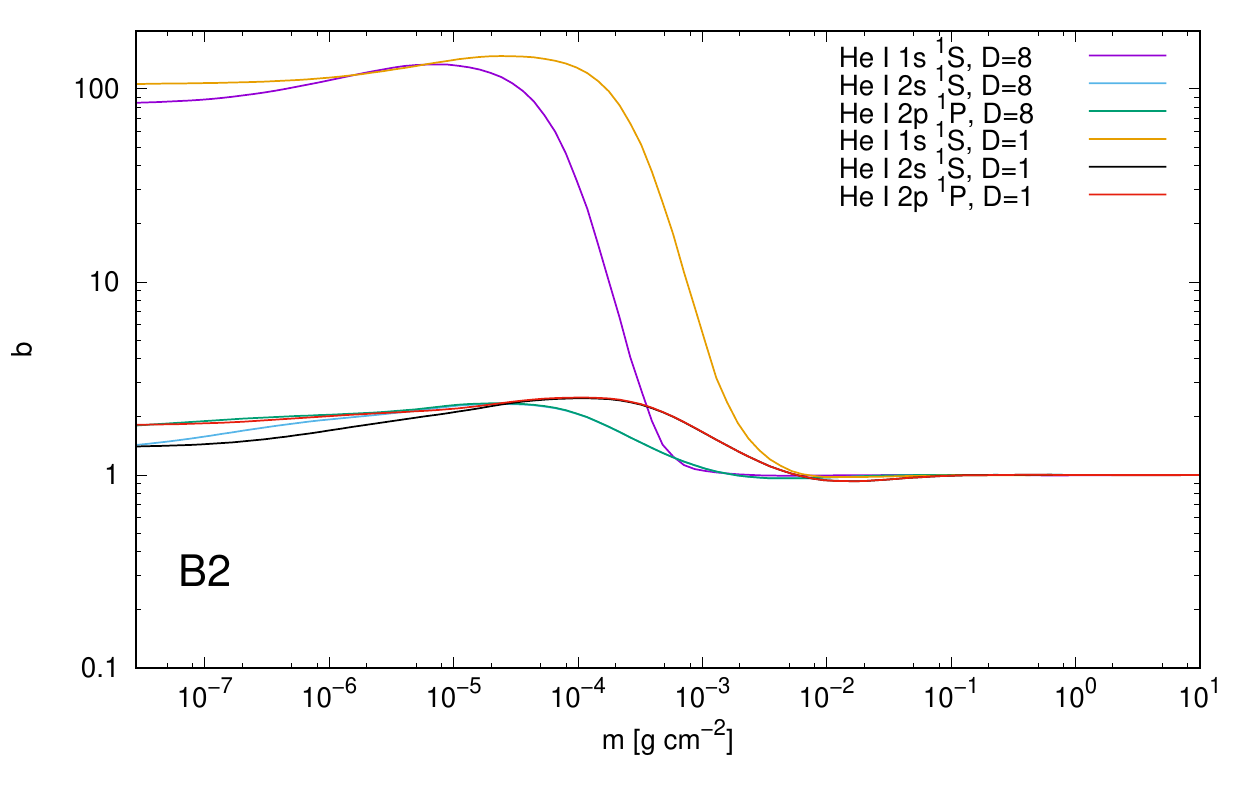}}
\resizebox{\hsize}{!}{\includegraphics{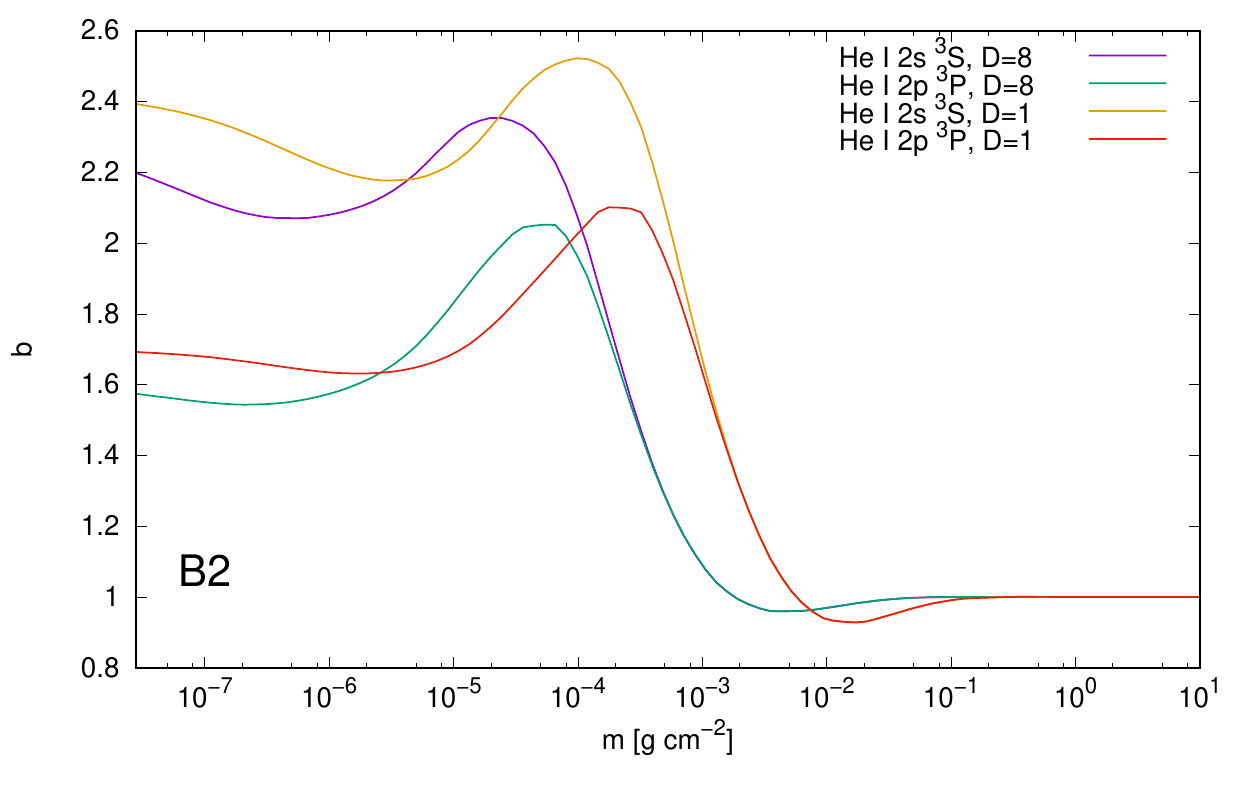}}
\resizebox{\hsize}{!}{\includegraphics{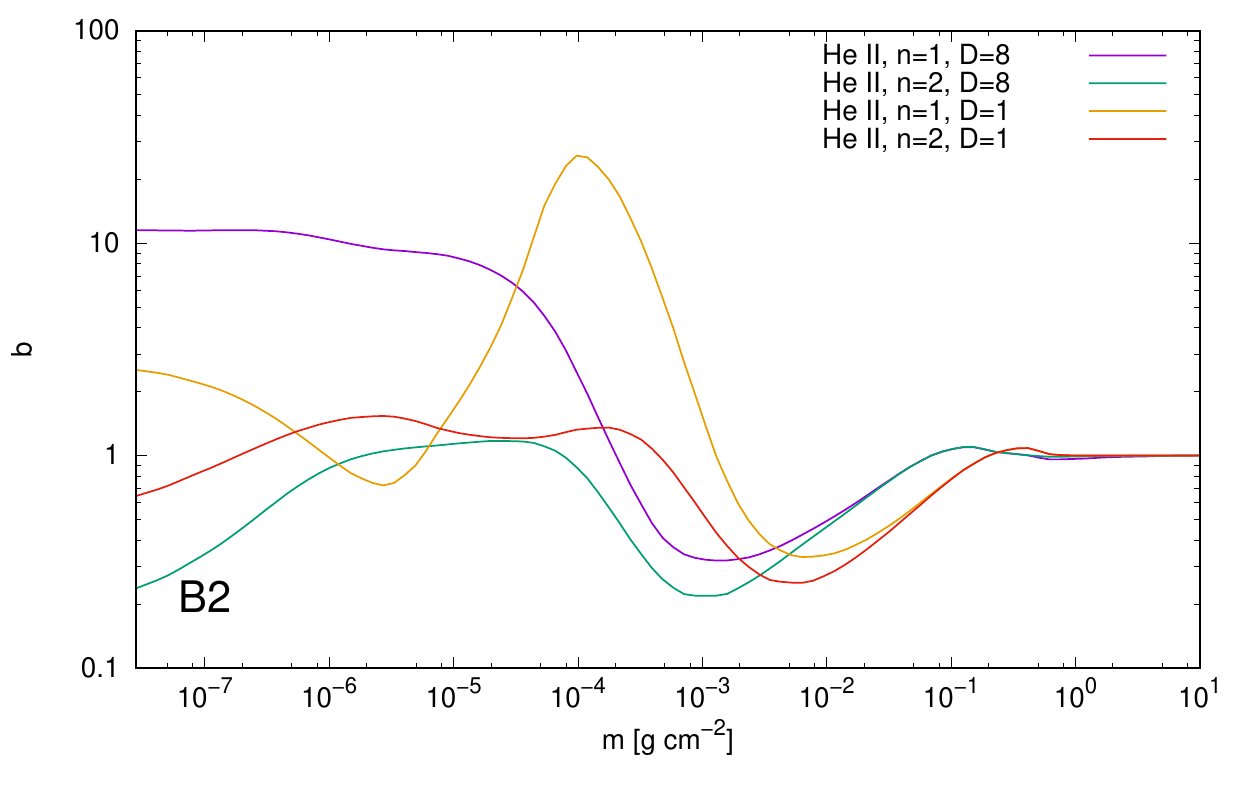}}
\caption{Same as Fig.\,\ref{55bfac}, but for the model set {\Bdva}.}
\label{B2bfac}
\end{figure}

\begin{figure}
\resizebox{\hsize}{!}{\includegraphics{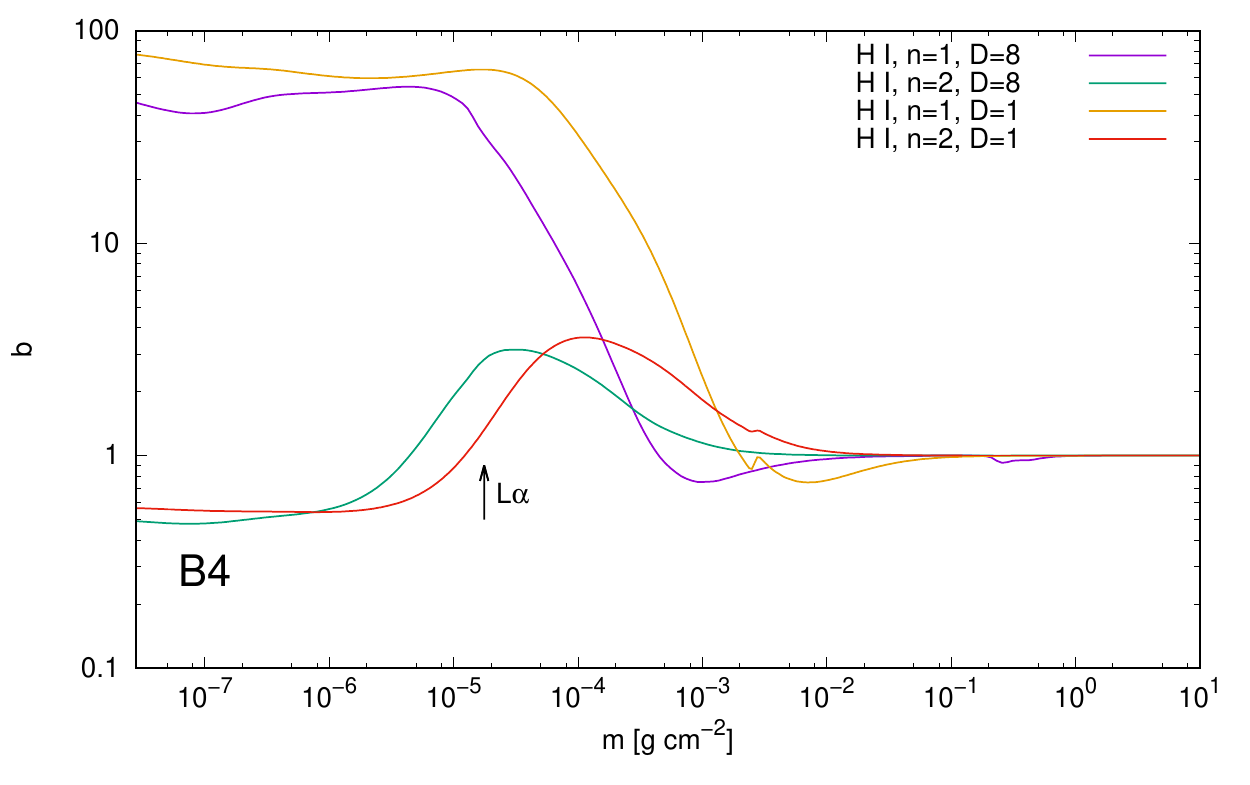}}
\resizebox{\hsize}{!}{\includegraphics{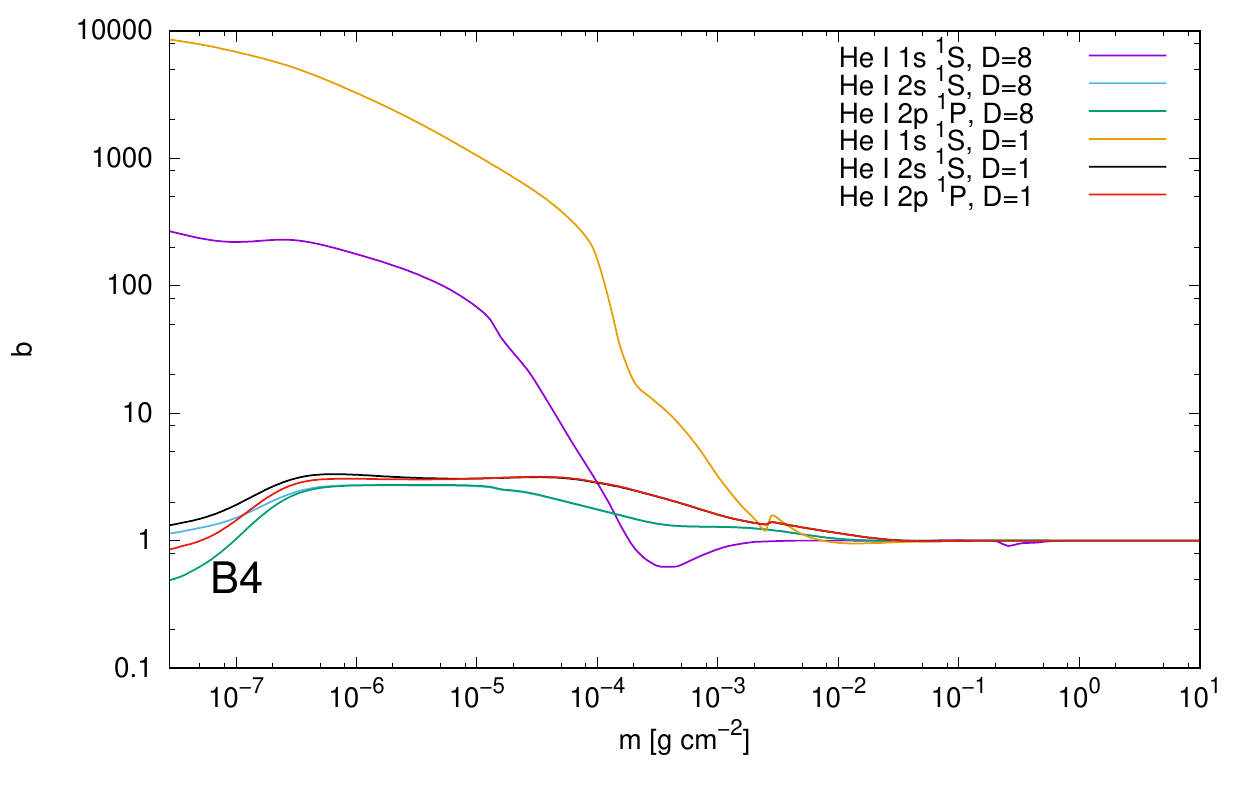}}
\resizebox{\hsize}{!}{\includegraphics{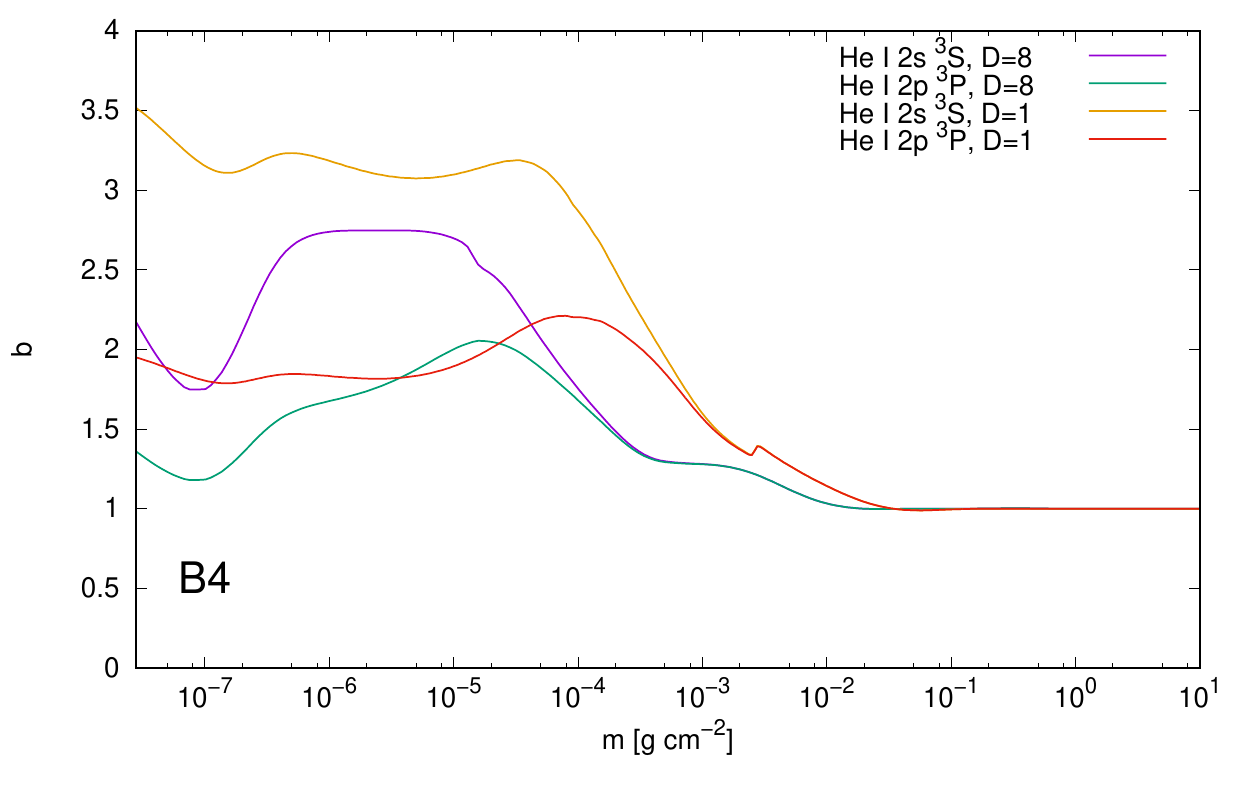}}
\caption{
Departure coefficients ($b$-factors) for selected atomic levels for clumping factors $\clfact=1$ and
$\clfact=8$ and for a set of model atmospheres {\Bctyri} (for basic global model parameters, see
Table\,\ref{parametry_modelu}).
From the top panel:
\ion{H}{i} $n=1$ and $n=2$;
\ion{He}{i} $1s\,\termls{1}{S}$, $2s\,\termls{1}{S}$, and $2p\,\termls{1}{P}$;
\ion{He}{i} $2s\,\termls{3}{S}$ and $2p\,\termls{3}{P}.$
}

\label{B4bfac}
\end{figure}

\begin{figure}
\resizebox{\hsize}{!}{\includegraphics{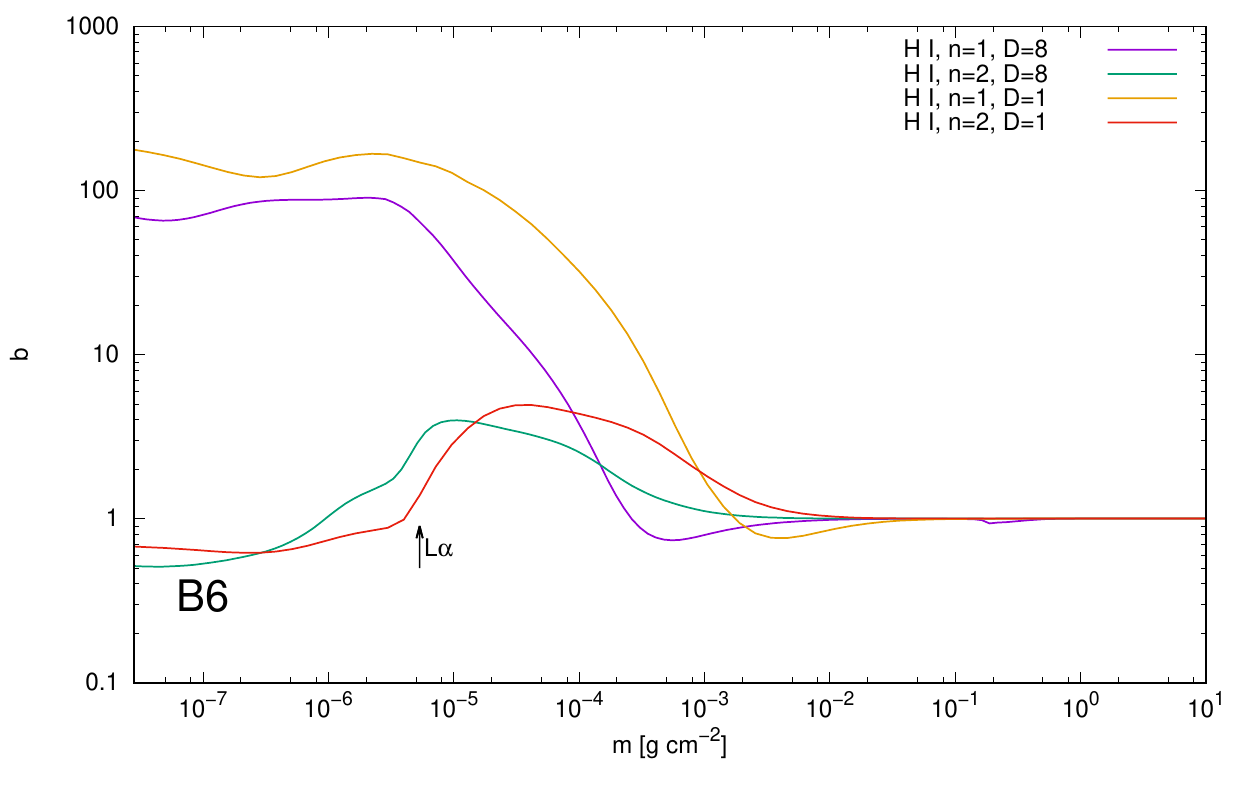}}
\resizebox{\hsize}{!}{\includegraphics{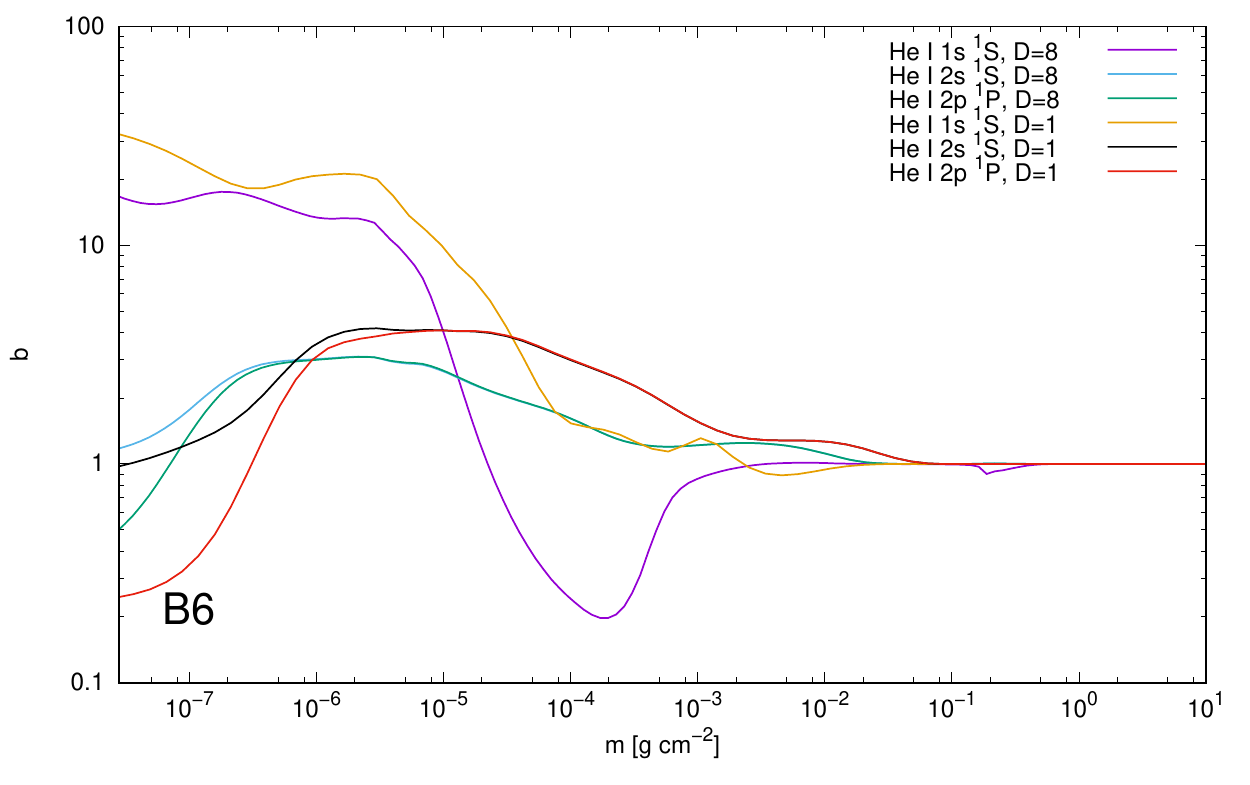}}
\resizebox{\hsize}{!}{\includegraphics{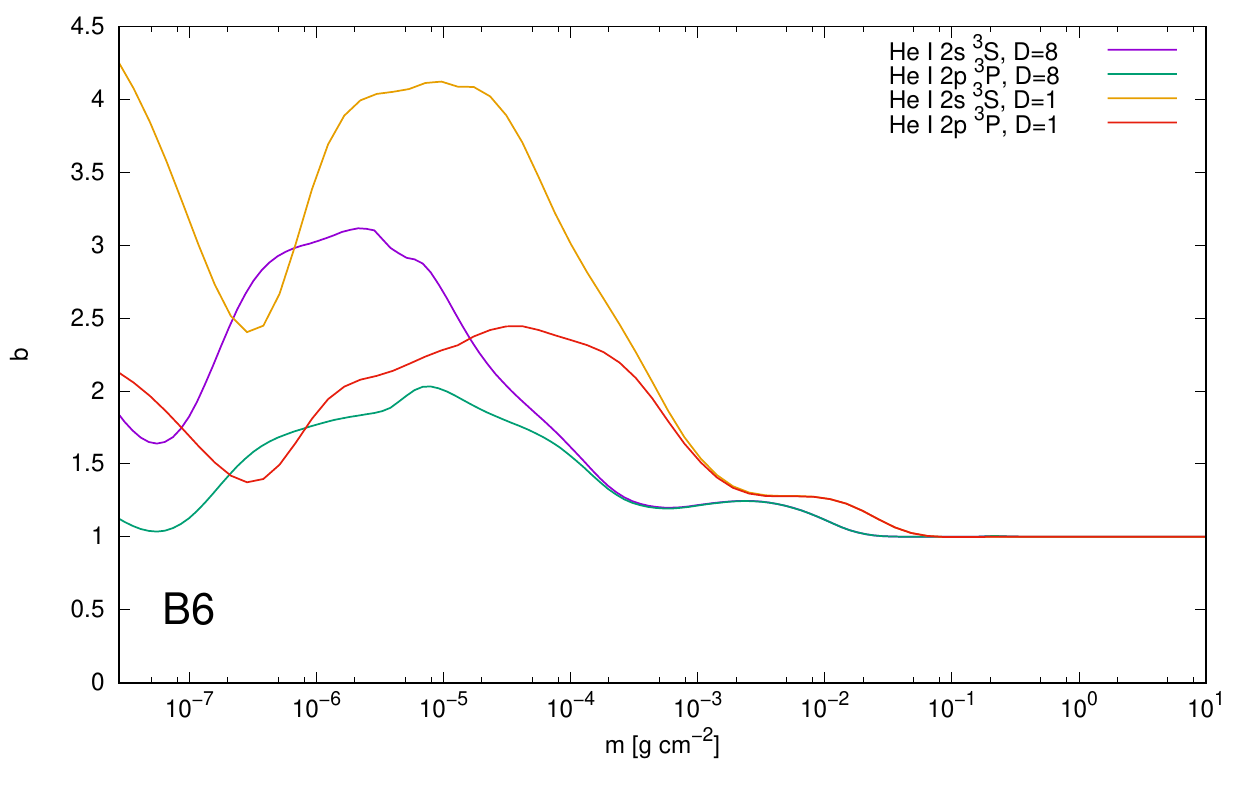}}
\caption{Same as Fig.\,\ref{B4bfac}, but for the model set {\Bsest}.}
\label{B6bfac}
\end{figure}

The NLTE effects in our model atmospheres cause departures from equilibrium values of atomic level
number densities $n_i$ and also strong ionisation shifts towards lower ionisation states for
optically thin parts of the atmospheres.
To eliminate the effect of the ionisation shift from departure coefficients, we used the common
definition of LTE populations in NLTE models, that is with respect to the ground level of the next
highest ion \citep[see][Eq.~14.228]{SA3}.
This means that we divided the {\bfactors} obtained from our models by the {\bfactor} of the ground
level of the next highest ion
\cite[current version of our code uses {\bfactors} after the original definition of][]{Menzel_1937}.
To give a more specific example, we divided the {\bfactor} of the \ion{He}{i} $1s\,\termls{1}{S}$
level by the {\bfactor} of the \ion{He}{ii} $n=1$ level.
We plotted these $b$-factors for the lowest levels of each ion included in the model calculation
(Figs. \ref{55bfac} to \ref{B6bfac}).

There is a trend of an increase of {\bfactors} with lower effective temperatures; this is
valid for both clumped and unclumped model atmospheres.
There is a remarkable similarity between {\bfactors} and their changes with clumping between the
hottest models of our sample (models {\sdpetpet}, {\Osest}, and {\Oosm}), although they are
calculated for significantly different stellar masses and radii.
The upper boundary of the line formation region can be roughly determined by the optical depth in
the line centre as the place where this optical depth is between 2/3 and 1.
This is indicated for the \ion{H}{i} and \ion{He}{ii} $\Lalpha$ lines by arrows in the upper and
bottom panels of Figs.~\ref{55bfac}, \ref{O6bfac}, and \ref{O8bfac}, respectively.
Interestingly, the `bump' in the depth dependence of the {\bfactor} of the $n=2$ level (which is
usually connected with the corresponding line formation region) moves outwards, while the region of
$\Lalpha$ formation (measured by the optical depth in the line centre) does not.
This is most probably caused by the effect of collisions, which are enhanced for a clumped
atmosphere.
Farther in the atmosphere at low optical depths, collisions are less frequent and radiative rates
dominate.
Enhanced {\bfactors} for the ground levels for clumped atmospheres there are caused by an increased
recombination rate due to clumping.
The effect of collisions can be observed also at the continuum formation region ($\tauross\approx
1$), for clumped atmospheres {\bfactors} start to deviate from the equilibrium value at higher $r$
(lower $m$).

For cooler models from our sample, only the \ion{H}{i} $\Lalpha$ line formation region is indicated
in Figs. \ref{B0bfac}, \ref{B2bfac}, \ref{B4bfac}, and \ref{B6bfac}.
For the case of the {\Bnula} model, the \ion{He}{ii} $\Lalpha$ line is optically thick throughout the
model atmosphere, while for the {\Bdva} model this optical thickness was assumed (the line was put
to the detailed radiative balance).
For models {\Bctyri} and {\Bsest}, the \ion{He}{ii} ion was not considered at all due to its low
abundance, which caused numerical problems.

For practically all stars where the ionised helium was considered, very large values of {\bfactors}
of the ground level of \ion{He}{ii} are noticeable.
This is a typical NLTE effect for strong resonance lines combined with a strong NLTE effect of an
active continuum \citep[which behaves like a resonance line; see][Chapter 14.6]{SA3}.
This leads to a massive overpopulation of the ground level and to a ionisation shift towards
\ion{He}{ii} (depopulation of \ion{He}{iii}) for $m\gtrsim 10^{-2}$.

Effects for \ion{He}{i} levels are more complex due to a different atomic structure.
For all models, there is a rise of the {\bfactor} for the ground level of \ion{He}{i} for low $m$.
For the hottest models this rise follows a significant decrease just above $\tauross\approx 1$.
Somewhat similar behaviour is seen for the metastable $2s\,\termls{1}{S}$ level; described effects
are weaker, however.
In addition, this level is closely coupled with the $2p\,\termls{1}{P}$ level below the formation
region of the corresponding line between these levels (20\,587\AA).
Above this region, standard behaviour (rise of the {\bfactor} of the lower level and decrease of the
{\bfactor} of the upper level) is observed.
The behaviour of the corresponding triplet levels ($2s\,\termls{3}{S}$ and $2p\,\termls{3}{P}$) is
qualitatively similar; collisional coupling between singlet and triplet levels plays a role in a
rise of departure coefficients of the displayed triplet levels.
Above the formation region of the 10\,830{\AA} line, a similar behaviour to that of singlet levels
is seen.
The behaviour of displayed \ion{He}{i} {\bfactors} is also influenced by interactions with other
levels taken into account (up to $n=4$ all fine structure levels in detail; from $n=5$ up to $n=9$
two mean levels for each quantum number $n$, one for singlets and one for triplets).

\section{Discussion}

Using our model atmosphere calculations, we studied the effect of clumping on the atmospheric
structure of {\jednad} static NLTE model atmospheres of hot dwarf stars.
Standard studies of clumping take the atmospheric structure (density and temperature) as given and
study the effects of clumping on radiative transfer and line formation.

Our calculations clearly show the effect of clumping on the vertical structure of {\jednad} NLTE
model atmospheres.
However, due to the {\trid} nature of clumping, inclusion of clumping in {\jednad} models must
suffer from approximations.
Here, we used the common and simplest approximate {\jednad} description of clumping, which assumes
that clumps are smaller than the photon mean-free path \citep{Hamann_Koesterke_1998} and means
that the clumps are optically thin.
This description is also referred to as the filling factor approach \citep{Hillier_1996,
Hillier_Miller_1999} or microclumping \citep[see][]{Oskinova_etal_2007} and uses the clumping factor
$\clfact$ (see Eq.\,\ref{clfactdef}) and the volume filling factor $\volfilfact$ (see
Eq.~\ref{rhomeandef}) for a quantitative description of clumping.
The advantage is that its solution lies in the radiative transfer equation with the same effort as
that of a horizontally homogeneous (smooth) atmosphere without clumps \citep{Hillier_2000}.

All resulting models are hotter in the region of continuum formation for larger clumping factors
$D$.
This backwarming effect is predominantly caused by increasing the local Rosseland mean opacity.

In our calculations, we assumed no clumping at significant depths.
Evidently, the single assumption of optically thin clumps is also acceptable at great optical depths
in the diffusion approximation region, and it causes a corresponding enhancement of the opacity.
However, the complete model of dense clumps (which must be very dense there) and a void interclump
medium is far from a realistic one at great optical depths, unless the matter is in a solid state
there.
To avoid this unrealistic situation, we used a simplifying assumption of no clumping ($D=1$) at
great depths ($\tauross\gtrsim 2/3$).
To check the effect of clumping at the dense part of the atmospheres, we also calculated
corresponding models with clumping at great depths.
Results of these calculations without the latter assumption, for a clumping factor that is constant
throughout the atmosphere, are shown in Fig.~\ref{sd55teplota1} for the case of the model
{\sdpetpet}.
This clumping model causes heating of the inner layers of the atmosphere, larger clumping factor $D$
causes higher temperatures, while the layers above $\tauross\gtrsim 2/3$ are almost the same as in
the case of no clumping in dense atmospheric parts (cf. Fig.~\ref{sd55teplota}).
Enhanced heating in the inner parts of the atmospheres causes larger emergent flux in the Lyman
series wavelength region (see Fig.~\ref{sd55flux1}).
We may conclude that clumping in the inner atmospheric parts enhances the effects of the emergent
flux described in Section~\ref{section_sed}; however, due to approximated {\jednad} treatment of
clumping, this conclusion has to be taken with care.
In any case, these effects underline the sensitivity of stellar atmospheres to different ways of
clumping treatment.

\begin{figure*}
\resizebox{0.5\hsize}{!}{\includegraphics{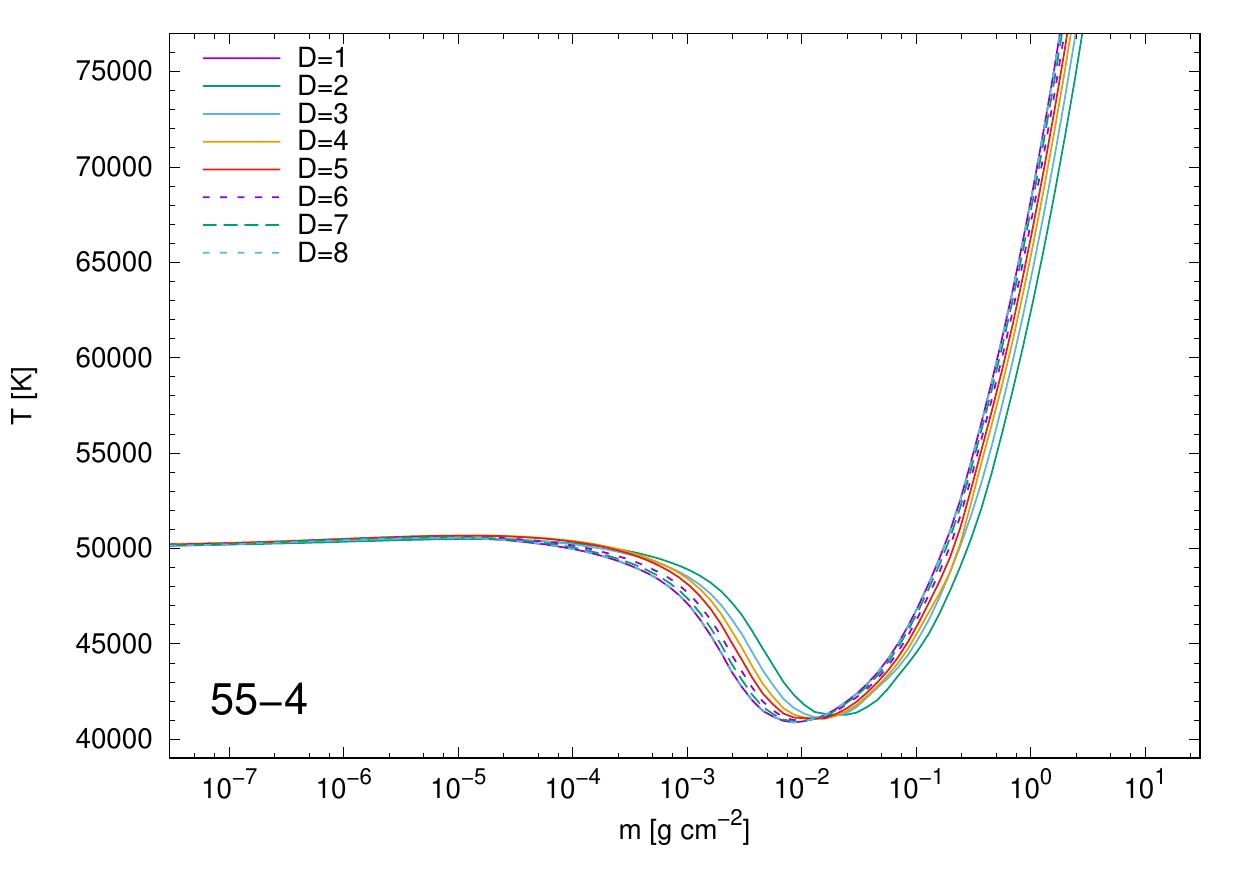}}
\caption{
Same as Fig.~\ref{sd55teplota}, but
with a depth-independent clumping factor.}
\label{sd55teplota1}
\end{figure*}

\begin{figure*}
\resizebox{\hsize}{!}{\includegraphics{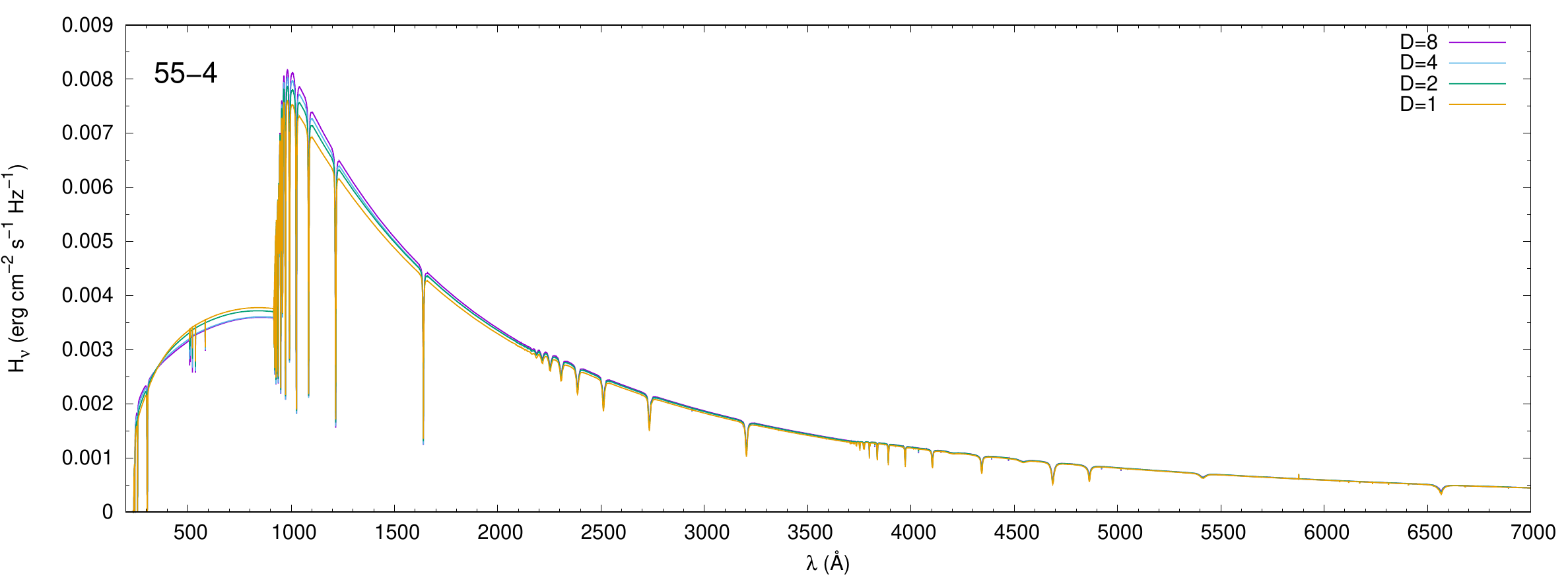}}
\caption{Same as Fig.~\ref{sd55flux}, but with a depth-independent clumping factor.}
\label{sd55flux1}
\end{figure*}

The really problematic region is the radiation formation region, around $\tau(\nu) \approx 1$, where
clumps may easily become larger than the photon mean-free path at a frequency $\nu$.
The only opacity influenced by the clumping factor in the optically thin approximation is the
free-free opacity, which is $\clfact$ times larger.
The bound-free and bound-bound opacities are the same as for the horizontally homogeneous (smooth)
atmosphere.
However, in reality the bound-bound and bound-free opacities in clumps may also become large, causing
the increase of the clump optical thickness $\tau_\clump(\nu)$ above 1 even for the case
when $\tau(\nu) \lesssim 1$ ($\tau(\nu)$ is evaluated using the mean opacity $\chimean$, see
Eq.\,(\ref{taunu}).
There, a {\trid} description should be preferred to include different properties of radiation
transfer through clumps (which may be optically thick) and the interclump medium (which is optically
thin, or in our case even void) in comparison to the smooth atmosphere, which may be optically thin.
Consequently, the optically thin clumping assumption underestimates the effect of clumping.

Another effect of clumping is more indirect and follows from the kinetic equilibrium equations and
changes in the emission coefficient.
This also includes the enhancement of collisional processes in clumps due to higher density there and
their dependence on the product $\nelec n_i$, and, in the case of collisional recombination, even on
$\nelec^2 n_i$ ($n_i$ is the corresponding atomic level number density).

As the clumping causes changes to the stellar emergent flux, it may affect basic stellar parameter
determination using model atmospheres.
Our NLTE models of clumped atmospheres of dwarf stars show higher flux in parts of the UV
region and lower flux in the visual and infrared regions.
This may cause the opposite effect to that of line blanketing.
Atmospheres may appear cooler in the visual region.
However, our set of parameters showed only small changes in the flux, at maximum several percent,
which lessens the significance of this effect.
Our data indicate that the effect of clumping is smaller for lower stellar effective temperatures;
to definitively confirm this statement, a more extended set of models would be necessary.
Furthermore, the effect on the emergent flux (which is opposite to the effect of line blanketing)
has to be verified using line-blanketed NLTE model atmospheres.

Besides continuum flux, some spectral lines are also affected.
However, we can not make a simple conclusion as to whether clumping causes the strengthening or
weakening of spectral lines, as we found both effects for the same models.
However, the changes are quite significant and likely affect the analysis; for example,
with regard to the abundance determination.
Unfortunately, the chemical composition of our models (only H+He) is too simple for real stars and
prevents us from making a direct comparison with observed stellar spectra and testing the expected
effects on abundance determination.

We also assumed microclumping (all clumps are optically thin).
This assumption, often used in the analysis of stellar winds, limits the clump properties and only
allows us to study basic effects of clumping.
As in the case of stellar winds, it is used mainly to simplify the problem.
On the other hand, in reality some clumps may easily become optically thick, either due to the
enhancement of density or through a more subtle effect of relaxing the assumption of LTE.
A generalised description of clumping (which allows macroclumping: i.e. optically thick clumping)
developed by \cite{Sundqvist_etal_2014} and used in stellar wind calculations by
\cite{Sundqvist_Puls_2018} uses an additional free parameter $h$ (the porosity length).
We decided to avoid introducing an additional free parameter, the depth dependence of which is still
a matter of discussion, and study the basic clumping effects in {\jednad} static NLTE model
atmospheres.

There is, of course, an uncertainty which values of the clumping factor $D$ are the proper ones for
the case of static atmospheres.
The clumping factor $D$ describes the density contrast of the clumped medium, so the high $D$ also
means significant enhancement of density inside clumps, and there should be a physical reason for
this.
If the clumping originates due to the sub-photospheric convection \citep{Cantiello_etal_2009}, we
can expect density contrasts related to the contrasts within sub-photospheric convective cells.
Consequently, we do not expect this contrast to be too high, so our decision to limit our
calculations to clumping factors $D\le8$ is reasonable.
In any case, to account for clumping properly, {\trid} models are to be preferred.

\section{Conclusions}

We present the first {\jednad} static NLTE model atmospheres of dwarf stars that consistently
include the effect of optically thin clumping on the vertical structure of the stellar atmosphere,
in addition to its influence on the emerging radiation.
Our models show differences between clumped and unclumped model atmospheres, even for the case of a
simplifying assumption of clumps smaller than the photon mean-free path (optically thin clumping).
For the more general case of optically thick clumps, greater effects can be expected.
However, this case has to be solved using {\trid} modelling.

\begin{acknowledgements}
The authors thank to the anonymous referee for his/her valuable comments to the manuscript.
This research has made use of the NASA’s Astrophysics Data System Abstract Service.
This research was supported by a grant 18-05665S (GA\v{C}R).
The Astronomical Institute Ond\v{r}ejov is supported by the
project RVO:67985815.
\end{acknowledgements}

\nocite{Ivan70}

%

\renewcommand*\bac{Bull. Astron. Inst. Czechosl.}
\newcommand{\aspconf}{ASP Conference Series}
\newcommand{\aspcs}{ASP Conf. Ser.}

\bibliographystyle{aa} 
\bibliography{m5proc,m5,kubat} 
%
\end{document}